\documentclass[revtex4]{emulateapj}

\usepackage{amssymb}
\usepackage{amsmath}
\usepackage[]{graphicx}
\usepackage{enumerate}
\usepackage{subeqnarray}
\usepackage{cases}
\usepackage{mathrsfs,amssymb}

\tightenlines

\begin{document}

\title{Penetration of Cosmic Rays into Dense Molecular Clouds: Role of Diffuse Envelopes\protect\footnote{
\MakeLowercase{\MakeUppercase{T}his paper is dedicated to the memory of \MakeUppercase{P}rof.
\MakeUppercase{V}adim \MakeUppercase{T}sytovich.}}}

\author{A.~V.~Ivlev$^1$, V.~A.~Dogiel$^2$, D.~O.~Chernyshov$^{2,3,4,5}$, P.~Caselli$^1$, C.-M.~Ko$^5$ \& K.~S.~Cheng$^3$}
\email[e-mail:~]{ivlev@mpe.mpg.de}

\affil{$^1$Max-Planck-Institut f\"ur extraterrestrische Physik, 85748 Garching, Germany }

\affil{$^2$I.~E.~Tamm Theoretical Physics Division of P.~N.~Lebedev Institute of Physics, 119991 Moscow, Russia}

\affil{$^3$Department of Physics, University of Hong Kong, Pokfulam Road, Hong Kong, China}

\affil{$^4$Moscow Institute of Physics and Technology (State University), Dolgoprudny, 141707, Russia}

\affil{$^5$Institute of Astronomy,  National Central University, Zhongli Dist., Taoyuan City, Taiwan (R.O.C.)}

\begin{abstract}
A flux of cosmic rays (CRs) propagating through a diffuse ionized gas can excite MHD waves, thus generating magnetic
disturbances. We propose a generic model of CR penetration into molecular clouds through their diffuse envelopes, and
identify the leading physical processes controlling their transport on the way from a highly ionized interstellar medium to
a dense interior of the cloud. The model allows us to describe a transition between a free streaming of CRs and their
diffusive propagation, determined by the scattering on the self-generated disturbances. A self-consistent set of equations,
governing the diffusive transport regime in an envelope and the MHD turbulence generated by the modulated CR flux, is
essentially characterized by two dimensionless numbers. We demonstrate a remarkable mutual complementarity of different
mechanisms leading to the onset of the diffusive regime, which results in a universal energy spectrum of the modulated CRs.
In conclusion, we briefly discuss implications of our results for several fundamental astrophysical problems, such as the
spatial distribution of CRs in the Galaxy as well as the ionization, heating, and chemistry in dense molecular clouds.
\end{abstract}


\maketitle

\section{Introduction}
\label{intro}

Cosmic rays (CRs) represent a crucial ingredient for the dynamical and chemical evolution of interstellar clouds.
Interaction of CRs with molecular clouds is accompanied by various processes generating observable radiation signatures,
such as ionization of molecular hydrogen \citep[see, e.g.,][]{oka05,dalg06,indri12} and iron
\citep[e.g.,][]{dog98,dog11,tati12,yus,nobu15,kriv17}, as well as production of neutral pions whose decay generates gamma
rays in the GeV \citep[e.g.,][]{yang1,yang2,tib15} and TeV \citep[e.g.,][]{ahar06, abram16,abdal17} energy ranges. Being a
unique source of ionization in dark clouds, where the interstellar radiation cannot penetrate, CRs provide a partial
coupling of the gas to magnetic field lines, which could slow down or prevent further contraction of the cloud
\citep[e.g.,][]{shu87}. CRs are fundamental for the starting of astrochemistry, as they promote the formation of H$_3^+$
ions, which can easily donate a proton to elements such as C and O, and thus eventually form molecules containing elements
heavier than H \citep[e.g.,][]{yamamoto17}. Through the ionization of H$_2$ molecules and the consequent production of
secondary electrons, CRs are an important heating source of dark regions \citep[e.g.,][]{goldsmith01}. Their interaction
with H$_2$ can also result in molecular excitation, followed by fluorescence producing a tenuous UV field within dark clouds
and dense cores \citep[][]{cecchi92,shen04,ivlev15a}; this UV field can photodesorb molecules from the icy dust mantles and
help maintaining a non-negligible amount of heavy molecules (such as water) in the gas phase \citep[e.g.,][]{caselli12}.
Furthermore, CRs can directly impinge on dust grains and heat up the icy mantles, causing catastrophic explosions of these
mantles \citep[][]{leger85,ivlev15b} and activating the chemistry in solids \citep[][]{shingledecker17}. Finally, CRs play a
fundamental role in the charging of dust grains and the consequent dust coagulation \citep[][]{okuzumi09,ivlev15a,ivlev16},
particularly important for the formation of circumstellar disks \citep[e.g.,][]{zhao16} and planet formation in more evolved
protoplanetary disks \citep[e.g.,][]{testi14}.

One of the fundamental questions is how interstellar (IS) CRs penetrate into molecular clouds, i.e., what are the governing
mechanisms of this process and how does this affect the CR spectrum inside the clouds. The crucial point here is that the IS
spectrum may be significantly modified while traversing the outer diffuse envelope of a cloud, before reaching the cloud
interior.

There are, at least, three important factors which may critically affect the CR spectra inside the clouds:
\begin{enumerate}
\item The cloud structure is strongly nonuniform. Dense cloud cores with the gas density $n_{\rm g}= 10^4-
    10^7$~cm$^{-3}$ are surrounded by low-density envelopes with $n_{\rm g}=10-10^3$ cm$^{-3}$
    \citep[see][]{lis90,protheroe}. In the central molecular zone these envelopes occupy up to 30\% of the space
    \citep[see][]{oka05,indri12}.
\item It is known since a long time \citep[see][]{lerche67,kuls69} that a CR flux propagating through a plasma can
    excite MHD waves and, thus, create magnetic disturbances. A linear analysis \citep[e.g.,][]{dog85} suggests that the
    waves are expected to be excited near most of the molecular clouds. However, it is still an open question as to
    whether the resulting disturbances are essential \citep[see][]{skill76,cesar78} or not \citep[see][]{gabici15} for
    the CR penetration into the clouds.
\item CR energy losses in the envelope are determined by ionization, proton-proton collisions, and MHD-wave excitation
    \citep[see][]{skill76,padovani, padovani1,ivlev15a, schlick16}. A relative importance of these processes also needs
    to be carefully analyzed.
\end{enumerate}

Attempts to analyze a system of nonlinear equations describing the CR-wave interaction in molecular clouds were undertaken
in several publications \citep[see, e.g.,][]{skill76,cesar78,gabici15}. We notice however, that in all these cases the
analysis was based on relatively simple estimates rather than on the exact solution of the equations. Nevertheless,
\citet{skill76} showed that interactions of CRs with waves should lead to depletion of their density inside the clouds at
energies below $\sim100$~MeV. Later, \citet{cesar78} demonstrated that the depletion can be even stronger if the effect of
magnetic field compression is taken into account. In a recent paper, \citet{gabici15} estimated the flux velocity of CRs
penetrating into a cloud to be about the Alfven speed for all energies. (Below we will see that this estimate is correct
only for relatively low energies.) For the sake of completeness, one should also mention analysis of the CR-wave interaction
undertaken by \citet{dog94} for processes of CR escape from the Galaxy, and by \citet{recc16a,recc16b} to describe the
spatial distribution of Galactic CRs and the CR-driven Galactic winds. These problems are, however, clearly beyond the scope
of our paper.

The principal goal of the present paper is an attempt to formulate a self-consistent generic model of CR penetration into
molecular clouds through their diffuse envelopes. We identify the leading physical processes controlling the CR propagation
on the way from a highly ionized interstellar medium to a dense interior of the cloud. In our analysis we do not presume a
regime of CR propagation in the envelope, but instead derive it from the model. This allows us to reveal the mutual
interplay of the factors mentioned above, and thus to address a number of important specific questions, such as:
\begin{enumerate}
\item What is the regime of CR propagation in molecular cloud envelopes -- do CRs freely cross the envelope, or do they
    experience significant scattering by the self-generated MHD turbulence?
\item What characteristics of the interstellar CR spectrum and parameters of a diffuse envelope determine the
    propagation regime?
\item Do CRs lose a significant part of their energy by MHD-wave excitation in the envelope, or do regular losses due to
    interaction with gas dominate?
\item Can (some of) the above processes cause a strong self-modulation of the CR flux penetrating into a dense core?
\end{enumerate}

The paper is organized as follows: In Section~\ref{Eqs} we present a self-consistent set of equations, governing the
diffusive regime of CR transport in a molecular cloud envelope and the MHD turbulence generated by the modulated CR flux. In
Section~\ref{normal} we write the governing equations in the dimensionless form and show that the diffusive regime is
described by a single dimensionless number $\nu$ (wave damping rate), while a transition to the free-streaming regime is
characterized by the small parameter $\epsilon$ (ratio of the Alfven velocity to the speed of light). In Section~\ref{model}
we consider an idealized problem setup, where CRs propagate toward an ``absorbing wall'' and the energy losses due to their
interaction with gas are negligible. This allows us to determine basic conditions of the onset of the diffusion zone in the
cloud envelope, and to identify generic properties of the nonlinear CR diffusion. In Section~\ref{realistic} we study the
effect of gas losses on the diffusion and, in particular, on the magnitude of the modulated CR flux penetrating into the
cloud. Finally, in Section~\ref{discussion} we point out a remarkable mutual complementarity of different mechanisms leading
to the onset of the diffusive regime, which results in a universal energy spectrum of the modulated CRs. Implications of our
results for several fundamental astrophysical problems are briefly discussed.

\section{Governing equations}
\label{Eqs}

In weakly ionized cloud envelopes, where the gas density $n_{\rm g}$ typically does not exceed $\sim10^3$ cm$^{-3}$, the
strength of the magnetic field $B$ is practically independent of $n_{\rm g}$ \citep[and is of the order of $10~\mu$G,
see][]{crutch12}. For this reason, we do not consider effects of large-scale variations of $B$, which may be essential for
CR propagation in dense cloud cores \citep[e.g.,][]{cesar78, schlick08}. Also, since the Larmor radius of CRs with energies
relevant to our problem is much smaller than the spatial extent of a typical envelope, a stream of such rapidly gyrating CRs
is parallel to the magnetic field. Hence, the problem can be considered as one-dimensional, with the coordinate $z$ measured
along the field line.

A CR flux can effectively excite Alfven and fast magnetosonic waves in a cold magnetized plasma. Low-frequency disturbances
of the magnetic field associated with these waves can, in turn, effectively scatter CRs. The maximum growth rate is achieved
for the waves propagating along the magnetic field in the direction of the CR flux. The growth rate is then the same for
both wave modes \citep[][]{kuls69}, propagating with the Alfven phase velocity,
\begin{equation*}
v_{\rm A}=\frac{B}{\sqrt{4\pi m_{\rm i}n_{\rm i}}}\,,
\end{equation*}
where $n_{\rm i}$ and $m_{\rm i}$ are the ion density and mean ion mass, respectively.

Let us introduce steady-state local distribution functions of CRs in the momentum and energy space, averaged over pitch
angle and denoted as $F(p,z)$ and $N(E,z)$, respectively. They are related to each other via
\begin{equation*}
4\pi p^2F(p,z)=vN(E,z)\equiv4\pi j(E,z),
\end{equation*}
where $j(E,z)$ is the so-called {\it CR energy spectrum}. The particle momentum as a function of the kinetic energy is
\begin{equation}\label{p(E)}
p(E)=c^{-1}\sqrt{E(E+2m_{\rm p}c^2)}\,,
\end{equation}
the physical velocity is $v(E)=p(E)c^2/(E+m_{\rm p}c^2)$. The local flux of CRs through a unit area and per unit energy
interval is defined as\footnote{The CR flux and hence the excited MHD waves propagate from right to left, as sketched below
in Figure~\ref{fig1}. Therefore, the minus sign is added in front of $S_{\rm free}$ and $v_{\rm A}N$ (note also that
$\partial N/\partial z\geq0$ in this case).}
\begin{equation}
S(E,z)\simeq-\min\left\{D\frac{\partial N}{\partial z}+v_{\rm A}N,~S_{\rm free}\right\}.
\label{smin}
\end{equation}
In such a definition, the flux continuously changes between the {\it diffusive regime} (first term; in what follows it is
referred to as the {\it modulated flux}), where the mean free path of CRs due to pitch-angle scattering on MHD turbulence is
sufficiently small, and the {\it free-streaming regime} (second term), where the scattering is negligible. For the former
regime, where the pitch-angle distribution is quasi-isotropic, the flux consists of the diffusion and advection parts
\citep[see, e.g.,][]{went74}, with $D(E,z)$ being the spatial diffusion coefficient of CRs. In turn, the magnitude of the
free-streaming flux,
\begin{equation}\label{sfree}
S_{\rm free}(E,z)=\langle\mu\rangle vN,
\end{equation}
is determined by average pitch angle of CRs in this regime, $\langle\mu\rangle$, which is generally not small. A discussion
of different free-streaming zones and estimates for the corresponding $\langle\mu\rangle$ is presented in
Appendix~\ref{<mu>}.

The steady-state CR flux is governed by the transport equation \citep[see, e.g.,][]{skill76,ber90}
\begin{equation}
\frac{\partial S}{\partial z}=-\frac{\partial }{dE}\left(\dot E_{\rm g}N\right),
\label{Diff_Eq}
\end{equation}
where $\dot E_{\rm g}(E)$ describes energy losses due to collisions with gas (``gas losses''). Here, we omit on purpose
``wave losses'', i.e., the term due to the adiabatic expansion of the magnetic disturbances associated with MHD waves. The
role of this term is discussed in Section~\ref{WL}, where we show that the wave losses are generally unimportant for our
problem. Furthermore, for waves propagating in one direction the mechanism of momentum diffusion (Fermi acceleration) does
not operate \citep[see, e.g.,][]{ber90}, and therefore the corresponding term is also not included in
Equation~(\ref{Diff_Eq}).

The diffusion coefficient of CRs \citep[][]{kuls69,ber90},
\begin{equation}\label{Diff}
D(E,z)=\frac{v^2}2\int_0^1d\mu\frac{1-\mu^2}{\nu_{\rm w}}\,,
\end{equation}
is determined by diffusion of their pitch angle $\mu$. The latter is characterized by the effective frequency of CR
scattering by MHD waves,
\begin{equation*}
\nu_{\rm w}(E,z,\mu)=2\pi^2\Omega_B(E)\frac{k_{\rm res}W(k_{\rm res},z)}{B^2}\,,
\end{equation*}
where $W(k,z)$ is the total spectral energy density of MHD waves, as discussed below, and $\Omega_B=(m_{\rm p}v/p)\Omega$ is
the gyrofrequency of a proton, expressed via gyrofrequency scale
\begin{equation*}
\Omega=\frac{eB}{m_{\rm p}c}\,.
\end{equation*}
Wavenumber $k_{\rm res}$ at a given energy is related to $\mu$ by a condition of the first-harmonic cyclotron resonance,
\begin{equation}\label{k_res}
|\mu| vk_{\rm res}=\Omega_B\,,
\end{equation}
or, equivalently, $|\mu| pk_{\rm res}=m_{\rm p}\Omega$. This condition assumes that $v$ is much larger than $v_{\rm A}$,
which sets a lower bound of $\sim\frac12m_{\rm p}v_{\rm A}^2$ for the kinetic energy of CRs in our consideration.

To identify generic effects of self-generated turbulence in weakly ionized envelopes, we assume no other sources of
turbulence and therefore no pre-existing MHD waves. The latter assumption is reasonable since, in the absence of internal
sources, such waves in a typical envelope experience relatively strong damping and therefore can be neglected compared to
the self-excited waves. The spectral energy density $W(k,z)$ for each wave mode is governed by a wave equation, including
dominant processes of excitation, damping, transport, as well as of nonlinear wave interaction. We employ the following
steady-state equation \citep[][]{laga83,norman96,ptus06}:
\begin{equation}
v_{\rm A}\frac{\partial W}{\partial z}+\frac{\partial }{\partial k}\left(\frac{kW}{T_{\rm NL}}\right)
=2(\gamma_{\rm CR}-\nu_{\rm damp})W.
\label{Wave_Eq}
\end{equation}

A nonlinear interaction of waves, leading to their cascading to larger $k$, is described in Equation~(\ref{Wave_Eq}) with
the simplest phenomenological model characterized by the cascade timescale $T_{\rm NL}$ \citep[][]{ptus06}. For the
Iroshnikov-Kraichnan cascade\footnote{In the following we demonstrate that the modulated CR flux is insensitive to the
particular model of cascade.} \citep{irosh64,kraich65} of acoustic MHD waves in an incompressible plasma, the timescale can
be evaluated as the characteristic time of ``collisions'' between oppositely traveling wave packets, $\sim(kv_{\rm
A})^{-1}$, multiplied by the number of collisions needed to accumulate a large distortion of the packets, $\sim m_{\rm
i}n_{\rm i} v_{\rm A}^2/(kW)$ \citep{gold97}. This yields
\begin{equation}
T_{\rm NL}^{-1}(k)=C_{\rm NL}\frac{k^2W(k)}{m_{\rm i}n_{\rm i} v_{\rm A}}\,,
\end{equation}
where $C_{\rm NL}\sim 1$ is an unknown constant. We assume $T_{\rm NL}$ to be the same for the excited MHD modes
\citep{gold97}, and then Equation~(\ref{Wave_Eq}) can be employed to describe the total spectral density of MHD waves.

The wave damping rate $\nu_{\rm damp}$ due to ion collisions with gas is proportional to the ratio $m_{\rm g}/m_{\rm i}$ of
the mean mass of a gas particle to the mean ion mass,
\begin{equation*}
\nu_{\rm damp}\simeq\frac12\frac{m_{\rm g}}{m_{\rm i}}\nu_{\rm g}\,.
\end{equation*}
It is determined by the momentum-transfer cross section of ion-gas collisions (averaged over velocities), $\nu_{\rm
g}=\langle\sigma v\rangle_{\rm ig} n_{\rm g}$. We recall that waves can only be sustained when their frequency exceeds the
damping rate, so for MHD waves the wavenumber should exceed the value of $\sim\nu_{\rm damp}/v_{\rm A}$ \citep{kuls69}. With
the resonance condition~(\ref{k_res}), this implies the upper limit on the energy of CRs that can contribute to the wave
excitation, $E\lesssim eBv_{\rm A}/\nu_{\rm damp}$. For typical conditions in diffuse envelopes ($n_{\rm
g}\sim100$~cm$^{-3}$, $B\sim10-100~\mu$G) we obtain the energy limit $\sim1-100$~TeV. This limitation does not affect the
results presented below, as the relevant energies turn out to be much smaller.

Finally, $\gamma_{\rm CR}$ is the (amplitude) growth rate of MHD waves excited by streaming CRs. These waves propagate along
the magnetic field in the same direction as the CR flux (to the left in Figure~\ref{fig1}), and their growth rate is given
by the following general formula, both for clockwise and counter-clockwise polarization \citep[][]{went74,skilb,ber90}:
\begin{eqnarray}
\gamma_{\rm CR}(k,z)=-\pi^3\frac{e^2v_{\rm A}}{c^2}\int_{-1}^{1}d\mu\:(1-\mu^2)\hspace{2.5cm}\label{gamma}\\
\times\int_0^{\infty}dp\:p^2v\:\delta(|\mu|pk-m_{\rm p}\Omega) \left(\frac{\partial f}{\partial \mu}-\frac{v_{\rm A}}{v}
p\frac{\partial f}{\partial p}\right),\nonumber
\end{eqnarray}
where $v\gg v_{\rm A}$ is assumed. Here, $f(p,z,\mu)\equiv F(p,z)+\delta f(p,z,\mu)$ is the anisotropic distribution of CR
in the momentum space, with $\langle\delta f\rangle_{\rm \mu}=0$, and $\delta(x)$ is the Dirac delta function. In the
diffusive regime {\it and} for a weak anisotropy, $|\delta f|\ll F$, the combination of derivatives in
Equation~(\ref{gamma}) is approximately equal to $-(v/\nu_{\rm w})\partial F/\partial z$ (the contribution of the gas losses
is normally negligible here). Taking onto account Equation~(\ref{Diff}), we see that in this case $\gamma_{\rm CR}$ is
determined by the diffusion part of the modulated CR flux. In Sections~\ref{model} and \ref{realistic} we discuss mechanisms
leading to the occurrence of gradients in the CR density.

Following \citet{skilb}, we introduce an effective cosine of the pitch angle, $\mu=\mu_*~(>0)$, in resonance condition
(\ref{k_res}). This provides one-to-one relation between $k_{\rm res}$ and $E$, reducing Equation~(\ref{k_res}) to
\begin{equation}\label{k_res1}
k_{\rm res}(E)=\frac{m_{\rm p}\Omega}{\mu_*p(E)}\,.
\end{equation}
With this approximation, elemental integration in Equation~(\ref{Diff}) yields a simple expression for the diffusion
coefficient,
\begin{equation}\label{Diff1}
D(E,z) \simeq \frac{1}{6\pi^2\mu_*}\frac{vB^2}{k^2W}\,,
\end{equation}
with $k^2W$ evaluated for $k(E)$ from Equation~(\ref{k_res1}). Similarly, by substituting $|\mu|=\mu_*$ in the
delta-function in Equation~(\ref{gamma}) and performing the integration, we derive
\begin{equation}\label{gamma1}
\gamma_{\rm CR}(k,z) \simeq\pi^2 \frac{e^2v_{\rm A}}{m_{\rm p}c^2\Omega} pvD \frac{\partial N}{\partial z}\,,
\end{equation}
where the (energy-dependent) rhs is evaluated for $E(k)$ from Equation~(\ref{k_res1}). Thus, with
approximation~(\ref{k_res1}) the growth rate is exactly proportional to the diffusion part of the modulated flux.
Equation~(\ref{gamma1}) remains applicable also in the free-streaming regime, after replacing $D\partial N/\partial z$ with
difference $S_{\rm free}-v_{\rm A}N$.

It is noteworthy that, generally, from Equations~(\ref{Diff}) and (\ref{gamma}) it follows that $D$ is a functional of
$W^{-1}$, and $\gamma_{\rm CR}$ is a functional of $W^{-1}\partial N/\partial z$. Effectively, this implies dependence of
$\mu_*$ on $k$, which can only be deduced by solving the resulting set of integral equations~(\ref{Diff_Eq}) and
(\ref{Wave_Eq}). However, this fact may only slightly change energy scalings of the results derived below with
approximation~(\ref{k_res1}), and therefore should not affect our principal conclusions.

\subsection{Role of wave losses}
\label{WL}

In Equation~(\ref{Diff_Eq}) we omitted wave losses -- a term representing the conventional adiabatic contribution,
proportional to the velocity gradient of MHD disturbances \citep[see, e.g.,][]{ber90}. After simple algebra, this term (to
be added under the energy derivative on the rhs) can be written as
\begin{equation*}
\dot E_{\rm w}N= -\frac13\frac{du}{dz}pvN,
\end{equation*}
where $u=-v_{\rm A}$ is the velocity of the disturbances in the diffusive regime. We see that for our problem the adiabatic
losses only operate at the border between the diffusion and the free-streaming zones, changing the CR flux by a value of
$\sim v_{\rm A}N$, i.e., of the order of the advection part in Equation~(\ref{smin}). Thus, the wave losses merely lead to a
renormalization of the advection.

In Sections~\ref{model} and \ref{realistic} we demonstrate that the advection part of the modulated flux can usually be
neglected for realistic conditions. Therefore, the wave losses are not expected to noticeably modify our results.

\section{Dimensionless units and dependence on physical parameters}
\label{normal}

To write governing equations (\ref{Diff_Eq}) and (\ref{Wave_Eq}) in a dimensionless form, we use the following normalization
of $E$, $k$, and $p$:
\begin{equation}\label{Ekp}
\tilde E=\frac{E}{m_{\rm p}c^2}\,,\quad \tilde k=\mu_*\frac{ck}{\Omega}=\frac1{\tilde p}=\frac1{\sqrt{\tilde E(\tilde E+2)}}\,,
\end{equation}
which naturally follows from Equations (\ref{p(E)}) and (\ref{k_res1}). In some cases it is also practical to utilize the
normalized physical velocity,
\begin{equation*}
\tilde v=\frac{\sqrt{\tilde E(\tilde E+2)}}{\tilde E+1}\,.
\end{equation*}
For brevity, we may use either of these variables to present results below.

Next, we introduce dimensionless CR spectrum,
\begin{equation*}
\tilde j=\frac{vN}{4\pi j_*}\,,
\end{equation*}
normalized by the characteristic value of the IS spectrum, $j_*=j_{\rm IS}(E=m_{\rm p}c^2)$. Now, in order to eliminate
coefficients in CR flux~(\ref{smin}) for the diffusive regime and, simultaneously, in wave equation~(\ref{Wave_Eq}), we
introduce dimensionless wave energy density $\tilde W=W/W_*$ and coordinate $\tilde z=z/z_*$, normalized by
\begin{equation*}
W_*=\frac{2\pi^2\mu_*^2}{C_{\rm NL}}\frac{m_{\rm p}^2c^3v_{\rm A}j_*}{\Omega}
\end{equation*}
and
\begin{equation}\label{z*}
z_*=\frac{C_{\rm NL}}{3\pi^3\mu_*}\frac{m_{\rm i}n_{\rm i}}{m_{\rm p}^2\Omega j_*}\,.
\end{equation}
Then Equations~(\ref{Diff_Eq}) and (\ref{Wave_Eq}) are reduced to
\begin{equation}\label{Diff_norm}
\frac{\partial \tilde S}{\partial\tilde z}=-\frac{\partial}{\partial \tilde p}\left(\tilde L_{\rm g}\tilde j\right),
\end{equation}
\begin{equation}\label{Wave_norm}
\tilde k^{3/2}\frac{\partial}{\partial \tilde k}\left(\tilde k^{3/2}\tilde W\right)
=\frac{\tilde D}{2\tilde k}\frac{\partial \tilde j}{\partial \tilde z}-\nu,
\end{equation}
where $\tilde L_{\rm g}$ and $\nu$ are dimensionless gas loss function and gas damping rate, respectively (both defined
later in this Section), while
\begin{equation}\label{D_norm}
\tilde D=\frac{\tilde v\tilde p^2}{\tilde W}\,,
\end{equation}
is the normalized diffusion coefficient. Dimensionless CR flux, $\tilde S=-\tilde v S/(4\pi j_*\epsilon)$, becomes
\begin{equation}
\tilde S=\min\left\{\tilde D\frac{\partial \tilde j}{\partial \tilde z}+\tilde j,~\tilde S_{\rm free}\right\},
\label{smin_norm}
\end{equation}
where the free-streaming term is
\begin{equation}\label{sfree_norm}
\tilde S_{\rm free}=\frac{\langle\mu\rangle}{\epsilon}\tilde v\tilde j.
\end{equation}
With the used normalization, the flux of free-streaming CRs is inversely proportional to the small parameter
\begin{equation}\label{epsilon}
\epsilon=\frac{v_{\rm A}}{c}\,,
\end{equation}
which is a measure of the contrast between the characteristic flux velocities in the two regimes (typically, $\epsilon\sim
10^{-3}-10^{-4}$). Note that in the transport equation~(\ref{Wave_norm}) we dropped the term $\sim\epsilon \tilde
W^{-1}\partial \tilde W/\partial \tilde z$ representing advection: Based on results of Section~\ref{approximate}, it is of
the order of $\epsilon\nu$ and therefore is negligible compared to the rhs.

The gas losses can be conveniently expressed in terms of the loss function $L_{\rm g}(E)=-\dot E_{\rm g}/n_{\rm g}v$, which
is a universal function of energy only (for a given gas composition). In the normalized form, it is
\begin{equation}\label{tilde_L}
\tilde L_{\rm g}=\frac1{\epsilon}\frac{n_{\rm g}z_*L_{\rm g}}{ m_{\rm p}c^2}\,.
\end{equation}
In the free-streaming regime, where $W\simeq0$, the small parameter $\epsilon$ cancels out in Equation~(\ref{Diff_norm}) and
CR transport naturally becomes independent of $v_{\rm A}$. Upon transition to the diffusive regime, the effective loss rate
is increased by a factor of $\epsilon^{-1}$, reflecting the corresponding increase of the distance traversed by self-trapped
CRs.

Thus, with the used normalization, the {\it only} dimensionless number entering governing Equations~(\ref{Diff_norm}) and
(\ref{Wave_norm}) (for a given loss function $L_{\rm g}$) is the damping rate
\begin{equation}\label{nu}
\nu=\frac{3\pi\mu_*}{4C_{\rm NL}}\frac{m_{\rm g}z_*\nu_{\rm g}}{m_{\rm i}c}\,,
\end{equation}
while the small parameter $\epsilon$ characterizes a transition between the diffusive and free-streaming
regimes.\footnote{For simplicity, the tilde sign over the dimensionless parameters $\nu$ and $\epsilon$ is omitted.}

The scaling dependence of $\nu$ and $\epsilon$ on the physical parameters is given by the following general expressions:
\begin{align}
\nu& = 8.7 \left(\frac{m_{\rm g}/m_{\rm p}}{2.3}\right)
    \left(\frac{j_* m_{\rm p} c^2}{1.3~\mbox{cm}^{-2}\mbox{s}^{-1}\mbox{sr}^{-1}}\right)^{-1} \label{scale_nu}\\
	&\times\left(\frac{n_{\rm i}/n_{\rm g}}{3\times 10^{-4}}\right)\left(\frac{n_{\rm g}}{100~\mbox{cm}^{-3}}\right)^2
    \left(\frac{B}{0.1~\mbox{mG}}\right)^{-1},\nonumber \\
\epsilon&= 1.2\times10^{-3} \left(\frac{m_{\rm i}/m_{\rm p}}{12}\right)^{-1/2}\label{scale_epsilon}\\
	&\times \left(\frac{n_{\rm i}/n_{\rm g}}{3\times 10^{-4}}\right)^{-1/2}\left(\frac{n_{\rm g}}{100~\mbox{cm}^{-3}}\right)^{-1/2}
    \left(\frac{B}{0.1~\mbox{mG}}\right).\nonumber
\end{align}
To give results in absolute units, we also use the normalization length,
\begin{align}
z_*&= 2.8 \times 10^{18}\frac{C_{\rm NL}}{\mu_*}\left(\frac{m_{\rm i}/m_{\rm p}}{12}\right)
    \left(\frac{j_* m_{\rm p} c^2}{1.3~\mbox{cm}^{-2}\mbox{s}^{-1}\mbox{sr}^{-1}}\right)^{-1}  \nonumber \\
	&\times\left(\frac{n_{\rm i}/n_{\rm g}}{3\times 10^{-4}}\right)\left(\frac{n_{\rm g}}{100~\mbox{cm}^{-3}}\right)
	\left(\frac{B}{0.1~\mbox{mG}}\right)^{-1}~\mbox{cm}. \nonumber
\end{align}

The illustrative numerical results presented in Sections~\ref{model} and \ref{realistic} are obtained by varying density of
gas $n_{\rm g}$. For simplicity, it is assumed that hydrogen is in molecular form and carbon photoionization by IS radiation
field is the main source of charged species \citep[see, e.g.,][]{oka06}. Hence, $m_{\rm g}/m_{\rm p}\simeq2.3$, $m_{\rm
i}/m_{\rm p}=12$, and $n_{\rm i}/n_{\rm g}\simeq 4\times 10^{-4}$, adopting the solar chemical composition with ionized
carbon. The magnetic field is set to $B = 100~\mu$G, in order to increase the magnitude of $\epsilon$ (which improves
convergence of the numerical scheme). For the ion-gas collisions we use $\langle\sigma v\rangle_{\rm ig} \simeq 2.1 \times
10^{-9}$~cm$^3$/s, corresponding to molecular hydrogen at a temperature of 100~K \citep[see, e.g.,][]{kuls69}. Finally, we
set $C_{\rm NL}=\mu_*=1$ and employ the following model spectrum for interstellar CRs \citep[][]{ivlev15a}:
\begin{equation}\label{IS_spectrum}
j_{\rm IS}(E)=\frac{1.4\times10^{-9}\tilde E^{-0.8}}{(0.55+\tilde E)^{1.9}}~~{\rm eV^{-1}cm^{-2}s^{-1}sr^{-1}}.
\end{equation}
With these physical parameters, $\nu$ and $\epsilon$ are related via
\begin{equation*}
\epsilon\nu^{1/4} = 1.7\times 10^{-3},
\end{equation*}
and below we only indicate the value of $\nu$.

In Appendix~\ref{numerical} we describe the algorithm to solve Equations~(\ref{Diff_norm}) and (\ref{Wave_norm})
numerically, and also give the gas loss function $L_{\rm g}(E)$ used to obtain numerical results presented in
Section~\ref{realistic}.

\section{A model problem: Absorbing wall}
\label{model}

\begin{figure}\centering
\includegraphics[width=0.9\columnwidth,clip=]{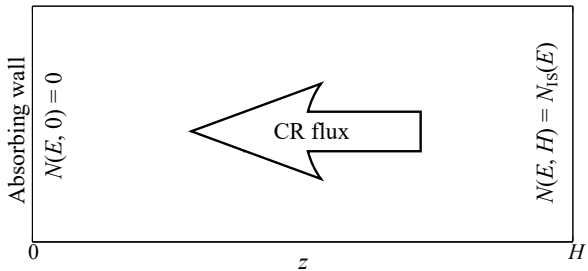}
\caption{Idealized problem setup with no gas losses. An absorbing wall is located at $z=0$, where the CR density
is set equal to zero. The incident IS flux propagates to the left, at the outer boundary $z=H$ the CR density is equal to
the IS value.} \label{fig1}
\end{figure}

We start with an idealized problem setup sketched in Figure~\ref{fig1}, and consider propagation of CRs toward an
``absorbing wall'' (which mimics a dense interior of a molecular cloud). The CR flux generates MHD turbulence upstream from
the wall (located at $z=0$), implying diffusive regime for CR propagation. Therefore, one can set $N(E,0)=0$ as the standard
boundary condition for the diffusion equation at an absorbing wall.\footnote{In fact, the CR density remains finite in the
diffusive regime: it is determined from the equality of the modulated and free-streaming fluxes in Equation~(\ref{smin}),
i.e., from condition $S=-S_{\rm free}$.} At the outer envelope boundary (located at $z=H$) the CR density is given by the
interstellar value, $N(E,H)=N_{\rm IS}(E)$. The principal aim of this simplified consideration is to identify generic
properties of nonlinear CR propagation, self-consistently described by the transport and wave equations discussed above.

We start with a case where the gas losses are unimportant, so the rhs of Equation~(\ref{Diff_norm}) can be set equal to
zero. Then the transport equation in the diffusive regime has a straightforward solution,
\begin{equation}\label{j_diff}
\frac{j(E,z)}{j_{\rm IS}(E)}=\frac{N(E,z)}{N_{\rm IS}(E)}=\frac{1-e^{-\eta(E,z)}}{1-e^{-\eta(E,H)}}\,,
\end{equation}
determined by ``diffusion depth''
\begin{equation}\label{zeta}
\eta(E,z)= \int_0^{\tilde z}\frac{dx}{\tilde D(E,x)}\equiv v_{\rm A}\int_0^z\frac{dx}{D(E,x)}\,.
\end{equation}
The magnitude of the resulting modulated flux~(\ref{smin}) is
\begin{equation}\label{flux}
S(E)=\frac{v_{\rm A}N_{\rm IS}(E)}{1-e^{-\eta(E,H)}}\,,
\end{equation}
(hereafter, we omit the minus sign in front of $S$). By virtue of Equation~(\ref{Ekp}) the solution can also be presented as
a function of $k$. One can see that $\eta$ is a measure of the relative importance of diffusion and advection in the
modulated CR flux: For $\eta\ll1$ Equation~(\ref{j_diff}) is reduced to the solution of the standard diffusion equation
($v_{\rm A}$ cancels out), for $\eta\gg1$ the CR density becomes constant and flux~(\ref{flux}) saturates at $v_{\rm
A}N_{\rm IS}$.

Below we show that the diffusive regime for given $E$ does not necessarily extend up to the outer envelope boundary, but may
terminate at the {\it outer border} of the diffusion zone $z_0(E)<H$, where $W\to0$. In this case, the free-streaming regime
with $N(E,z)=N_{\rm IS}(E)$ operates at $z>z_0$, and the solution does not depend on $H$.

By substituting Equation~(\ref{j_diff}) in Equation~(\ref{Wave_norm}) we derive the following wave equation for
self-consistent turbulent field in the diffusive regime:
\begin{equation}\label{W_SC}
\tilde k^{3/2}\frac{\partial}{\partial \tilde k}\left(\tilde k^{3/2} \tilde W\right)
=\frac{\tilde j_{\rm IS}(k)}{2\tilde k}\frac{e^{-\eta(k,z)}}{1-e^{-\eta_0(k)}}- \nu,
\end{equation}
where $\eta(k,z)$ is given by Equation~(\ref{zeta}) with $E(k)$ from Equation~(\ref{Ekp}),
\begin{equation*}
\eta(k,z)=\tilde k^2\sqrt{1+\tilde k^2}\int_0^{\tilde z}dx\:\tilde W(k,x),
\end{equation*}
and $\eta_0(k)=\eta(k,z_0)$. We recall that the excitation term in Equation~(\ref{W_SC}) is proportional to the diffusion
part of the modulated flux which, in turn, cannot exceed the flux of free streaming CRs. Then from
Equations~(\ref{smin_norm}) and (\ref{sfree_norm}) it follows that in the diffusive regime, with $j(E,z)$ from
Equation~(\ref{j_diff}), condition $\eta_0\gtrsim v_{\rm A}/v$ must always be fulfilled. This lower bound of $\eta_0$ (which
is a small number, since $v\gg v_{\rm A}$ is assumed) represents the necessary condition of applicability for the diffusion
approximation.

We notice that requirement
\begin{equation}\label{zeta_min}
\eta\gtrsim \frac{v_{\rm A}}{v}
\end{equation}
coincides with the condition that the mean free path of CRs, $\sim D/v$, is smaller than the inhomogeneity scale length,
$\sim N/|\partial N/\partial z|$, as one can easily derive from Equations~(\ref{j_diff}) and (\ref{zeta}); simultaneously,
this ensures that the velocity of the CR flux does not exceed the physical velocity. Therefore, we shall consider
inequality~(\ref{zeta_min}) as the sufficient condition of applicability of the diffusion approach. The resulting {\it inner
border} of the diffusion zone $z_{\rm min}(E)$ is determined from condition $\eta(E,z_{\rm min})\sim v_{\rm A}/v$.

The threshold energy $E_{\rm ex}$, below which CRs excite waves, can be readily derived from the balance of the growth rate
in the {\it free-streaming} regime and the damping rate. By replacing the diffusion flux on the rhs of
Equation~(\ref{Wave_norm}) with the free-streaming expression from Equation~(\ref{sfree_norm}), we obtain the following
equation:
\begin{equation}\label{threshold}
\frac{\tilde E_{\rm ex}+2}{\tilde E_{\rm ex}+1}\tilde E_{\rm ex}\tilde j_{\rm IS}(E_{\rm ex})
=\frac{2\epsilon \nu}{\langle\mu\rangle}\,,
\end{equation}
where $\langle\mu\rangle$ is the average pitch angle in the free-streaming zone I (see Appendix~\ref{<mu>} and
Figure~\ref{figA1} therein). For sufficiently steep, monotonic energy spectra, e.g., $\tilde j_{\rm IS}= \tilde E^{-\alpha}$
with $\alpha>1$, waves are excited if $E<E_{\rm ex}$; the threshold energy scales as
\begin{equation*}
E_{\rm ex}\propto\left(\frac{m_{\rm g}n_{\rm g}}{j_*}\sqrt{\frac{n_{\rm i}}{m_{\rm i}}}\:\right)^{-\frac1{\alpha-1}}.
\end{equation*}
Equation~(\ref{threshold}) also shows that CRs with $j_{\rm IS}\propto E^{-1}$ represent a {\it critical case}, where the
excitation occurs when the flux magnitude matches the damping threshold.

Numerical analysis shows that the magnitude of $W$ in the turbulent zone is typically high enough for the condition of the
diffusion approximation to be well fulfilled. Thus, it is reasonable to solve wave equation~(\ref{W_SC}) for $k>k_{\rm
ex}\equiv k(E_{\rm ex})$ with condition $W(k_{\rm ex},z)=0$. The solution in $(k,z)$ space is applicable for
$\eta(k,z)\gtrsim v_{\rm A}/v$, while outer turbulent border $z_0(k)$ is obtained from $W(k,z_0)=0$.

\subsection{Approximate solution}
\label{approximate}

One can obtain a simple approximate solution of Equation~(\ref{W_SC}), providing a fairly accurate and general description
of the turbulent regime. From the numerical integration performed for different values of $\nu$ we found that, as long as
$\eta_0\lesssim1$ and $\nu$ is not too small, the turbulent field can be reasonably approximated by a decreasing linear
function of coordinate (see Appendix~\ref{linear_W} and the figure therein),
\begin{equation}\label{field}
\tilde W(k,z)\simeq w(k)+w'(k)\tilde z,
\end{equation}
with $w'<0$, so the outer border of the diffusion zone is $\tilde z_0(k)=-w(k)/w'(k)$. Equation~(\ref{field}) breaks down
close to $k_{\rm ex}$, but this does not affect properties of the whole diffusion zone.

We first study the case of small diffusion depth, $\eta_0\lesssim1$, which allows us to expand the exponentials on the rhs
of Equation~(\ref{W_SC}). We retain only linear terms in the resulting $z$-polynomial and equate to zero the corresponding
coefficients, which gives us two equations for $w(k)$ and $w'(k)$. One equation yields
\begin{equation}\label{zeta_0}
\eta_0(k)=\frac{\tilde j_{\rm IS}(k)}{2\tilde k\nu}\,,
\end{equation}
which is simply a balance of the excitation and damping on the rhs of Equation~(\ref{W_SC}), written for small $\eta$; the
lhs, i.e., the cascade term for $w(k)$, is neglected here compared to $\nu$ -- this assumption is confirmed {\it a
posteriori}. The other equation leads to
\begin{equation*}
\frac{d}{d\tilde k}\left(\tilde k^{3/2}w'\right)=\nu \sqrt{\tilde k(1+\tilde k^2)}\:w'\tilde z_0\,,
\end{equation*}
showing that the cascade is essential for $w'(k)$. By combining Equation~(\ref{zeta_0}) with relation $\eta_0(k)=
-\frac12\tilde k^2 \sqrt{1+\tilde k^2}\:w'\tilde z_0^2$ and setting $w'(k_{\rm ex})=0$, we get the solution which can be
conveniently written as
\begin{equation}\label{w'}
2\sqrt{\frac{-\tilde k^{3/2}w'(k)}{\nu}}=\int_{\tilde k_{\rm ex}}^{\tilde k}dx\sqrt{\frac{\sqrt{1+x^2}\:
\tilde j_{\rm IS}(x)}{x^{7/2}}}\,.
\end{equation}
Then $\tilde z_0(k)$ is readily obtained by employing the above relation for $\eta_0(k)$, and $w(k)=-\tilde z_0(k)w'(k)$. We
note that a realistic IS spectrum, such as Equation~(\ref{IS_spectrum}), is a rather steeply increasing (decreasing)
function at small $\tilde k$ (large $\tilde E$). Therefore, if $\tilde k_{\rm ex}\lesssim1$, the integral in
Equation~(\ref{w'}) is dominated by larger $k$, i.e., the contribution of $k\simeq k_{\rm ex}$ vanishes asymptotically.

With this solution we can verify the simplifications/assumptions made to obtain it: First, we recall that the advection term
$\sim\epsilon \tilde W^{-1}\partial \tilde W/\partial \tilde z$ was dropped in Equation~(\ref{Wave_norm}). For $k\gg k_{\rm
ex}$ we get $\epsilon\tilde W^{-1}|\partial \tilde W/\partial \tilde z|\simeq\epsilon|w'|/w\sim\epsilon\nu\sqrt{1+\tilde
k^2}$, which is indeed small compared to $\nu$. Second, by substituting solution $w(k)\sim\tilde j_{\rm IS}(k)/\tilde k^3$
to the cascade term in the lhs of Equation~(\ref{W_SC}) we conclude that the latter is small compared to $\nu$ too, as long
as $\eta_0\lesssim1$.

Condition $\eta_0\lesssim1$ implies a certain upper limit on $k$, since $\eta_0(k)$ is an increasing function (for realistic
IS spectra). For larger $\eta_0$ (and $k$), numerical results indicate that spatial nonlinearity of the turbulent field
becomes significant (see Appendix~\ref{linear_W}). Nevertheless, Equation~(\ref{field}) still provides useful qualitative
description of the diffusion zone. For $\eta_0\gg1$, term $e^{-\eta_0}$ in Equation~(\ref{W_SC}) can be neglected. In this
case, to determine $w(k)$ and $w'(k)$ we write the resulting wave equation for $z=0$ and $z=z_0$. The former gives
\begin{equation}\label{w}
\tilde k^{3/2}\frac{d}{d\tilde k}\left(\tilde k^{3/2}w\right)=\frac{\tilde j_{\rm IS}(k)}{2\tilde k}-\nu,
\end{equation}
showing that excitation exceeds damping at larger $k$, so that now the cascade plays a crucial role. In the latter equation,
we neglect the term $\propto e^{-\eta_0}$ and, after simple transformation, obtain the following equation for $z_0(k)$:
\begin{equation}\label{z_0}
\frac{d\ln z_0}{d\tilde k}=-\frac{\nu}{\tilde k^3w(k)}\,.
\end{equation}
Equation~(\ref{w}) allows straightforward integration for given $j_{\rm IS}(k)$, and the derived $w(k)$ has to be matched
with that obtained from Equations~(\ref{w'}). By substituting the result in Equation~(\ref{z_0}) and integrating it, we get
$z_0(k)$ for large $\eta_0$.

\subsection{Diffusion zone}
\label{zone}

\begin{figure}\centering
\includegraphics[width=\columnwidth,clip=]{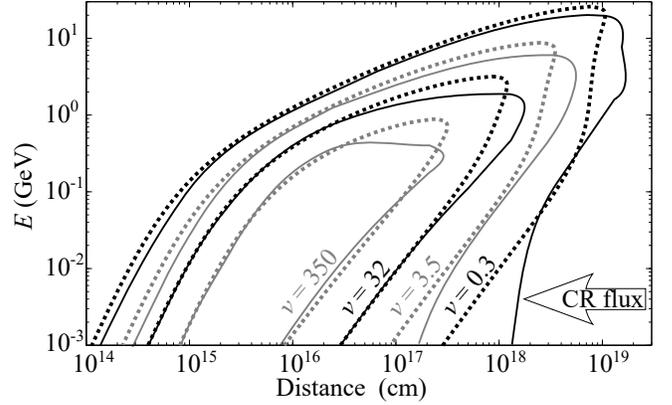}
\caption{CR diffusion zones: regions in $(E,z)$ plane within which the CR propagation is diffusive. The solid lines are the
numerically calculated borders, plotted for different values of $\nu$ (indicated) and $\epsilon\propto\nu^{-1/4}$ (see
Section~\ref{normal} for details). The dotted lines show analytical inner (left) and outer (right) borders, $z_{\rm min}(E)$
and $z_0(E)$, respectively, derived from solution~(\ref{w'}) for given $\nu$.} \label{fig2}
\end{figure}

Figure~\ref{fig2} illustrates the characteristic form of the diffusion zone in $(E,z)$ plane. The numerically calculated
diffusion border is plotted for several values of $\nu$ (solid lines). The right branch of each contour is the outer border
of the zone $z_0(E)$, approximately derived in Section~\ref{approximate}, while the left branch corresponds to inner border
$z_{\rm min}(E)$, determined by condition~(\ref{zeta_min}). The branches cross at the highest ``critical'' point $E\simeq
E_{\rm ex}(\nu)$, determined by Equation~(\ref{threshold}). The analytical curves $z_0(E)$ and $z_{\rm min}(E)$, obtained
from solution~(\ref{w'}) (dotted lines), demonstrate a good overall agreement with the numerical results. A stronger
deviation is observed toward the critical point, where the approximate solution breaks down. Also, at lower energies
analytical $z_0(E)$ deviates increasingly from the numerical curve when $\nu$ is small.

Using solution~(\ref{w'}), one can deduce how the shape of the diffusion zone depends on the form of the IS spectrum and the
main physical parameters. For $\tilde j_{\rm IS}(E)=\tilde E^{-\alpha}$ with $\alpha(E)$ determined by a model spectrum,
Equation~(\ref{IS_spectrum}) or analogous \citep[][]{ivlev15a}, it is practical to consider two limiting cases -- the
ultra-relativistic limit, where $\tilde k=1/\tilde E\ll1$, and the non-relativistic case, where $\tilde k=1/\sqrt{2\tilde
E}\gg1$. Equation~(\ref{w'}) yields the outer border, $\tilde z_0(E)\sim1/\nu$ for $\tilde E\gg1$ and $\tilde
z_0(E)\sim\sqrt{\tilde E}/\nu$ for $\tilde E\ll1$. Substituting a solution for $w(k)$ in condition $\eta(E,z_{\rm
min})\sim\epsilon/\tilde v$, we obtain the inner border, $\tilde z_{\rm min}(E)\sim\epsilon \tilde E^{\alpha-1}$ for $\tilde
E\gg1$ and $\tilde z_{\rm min}(E)\sim\epsilon \tilde E^{\alpha-1/2}$ for $\tilde E\ll1$. In absolute units, this gives the
following dependence on the physical parameters:
\begin{equation}\label{scaling_z1}
z_{\rm min}\propto \frac{\sqrt{m_{\rm i}n_{\rm i}}}{j_*}\,,\quad z_0\propto \frac{m_{\rm i}}{n_{\rm g}}\,.
\end{equation}

If $\eta_0\gtrsim1$, which corresponds to large $k$ and/or small $\nu$, solution~(\ref{w'}) is no longer applicable and the
turbulent field is qualitatively described by Equations~(\ref{w}) and (\ref{z_0}). The former yields $\tilde k^3
w(k)\sim\tilde j_{\rm IS}(k)$ for large $k$, and then from the latter equation we invoke that $z_0(k)$ asymptotically tends
to a constant value. This explains the behavior of numerically calculated $z_0(E)$ at lower $E$ and small $\nu$, seen in
Figure~\ref{fig2} for $\nu=0.3$ and 3.5.

The diffusion zone is formed when $z_{\rm min}(E)\lesssim z_0(E)$. Using the above estimates for the inner and outer
borders, we then arrive to a simple criterion of the diffusive regime, valid for all energies where $\eta_0(E)\lesssim1$:
\begin{equation}\label{zone_eq}
\epsilon\nu \tilde E^{\alpha-1}\lesssim1.
\end{equation}
Expectedly, this criterion is essentially equivalent to the excitation criterion~(\ref{threshold}) in the free-streaming
regime. Equation~(\ref{zone_eq}) shows that if $\alpha>1$ for any $E$, the diffusion zone shrinks monotonically with $\nu$
toward lower energies, until the basic resonance condition~(\ref{k_res}) becomes inapplicable at $v\lesssim v_{\rm A}$.
Current models of the IS spectra, such as Equation~(\ref{IS_spectrum}), suggest $\alpha<1$ for non-relativistic CRs. Then
the diffusion zone for sufficiently large $\nu$ becomes an isolated ``island'', and eventually disappears when product
$\epsilon\nu$ exceeds a certain maximum value $(\epsilon\nu)_{\rm max}\sim1$. The exact value of $(\epsilon\nu)_{\rm max}$
is derived from Equation~(\ref{threshold}) and corresponds to the maximum of its lhs; e.g., for IS
spectrum~(\ref{IS_spectrum}) the maximum is at $E\simeq60$~MeV, and $(\epsilon\nu)_{\rm max}\sim1$. Then from
Equations~(\ref{scale_nu}) and (\ref{scale_epsilon}) we obtain the maximum gas density $n_{\rm g}\sim3\times10^4$~cm$^{-3}$,
above which no turbulence can be excited by CRs with such energy spectrum.\footnote{We note that the obtained maximum gas
density is about the {\it average} density inside dense cores \citep[e.g.,][]{benson89}.} In Figure~\ref{fig2}, the
diffusion zone completely disappears at $\nu\sim3\times10^3$.

Figure~\ref{fig2} also indicates that, for very small $\nu$, the derived outer border $z_0(E)$ at higher energies may be
larger than the envelope size $H$. Then the diffusion zone is bound between $z_{\rm min}(E)$ and $H$, and the solution
obtained in Section~\ref{approximate} for $W(k,z)$ is modified. Nevertheless, as long as the resulting
$\eta(k,H)\equiv\eta_H$ is small, its value is determined from the same excitation-damping balance that leads to
Equation~(\ref{zeta_0}), and therefore $\eta_H$ is equal to the derived $\eta_0$. In this case, the condition of the
diffusion regime to operate is simply $z_{\rm min}(E)\lesssim H$.

\subsection{CR flux}
\label{CR_flux}

\begin{figure}\centering
\includegraphics[width=\columnwidth,clip=]{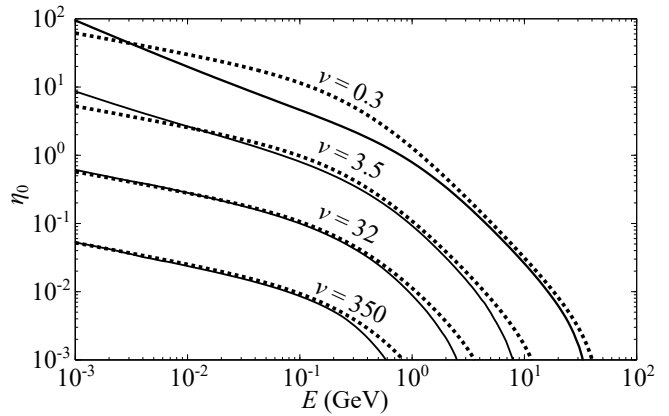}
\caption{``Diffusion depth'' $\eta_0(E)$, numerically calculated (solid lines) for the values of $\nu$ in Figure~\ref{fig2}.
The analytical dependence (dotted lines) given by Equation~(\ref{zeta_0}) provides good description for $\eta_0\lesssim1$.
Each curve tends to zero at $E=E_{\rm ex}(\nu)$ determined by Equation~(\ref{threshold}). } \label{fig3}
\end{figure}

>From Equation~(\ref{flux}) it follows that the value of diffusion depth $\eta_0$ (or $\eta_H$) completely determines the CR
flux penetrating the cloud. Figure~\ref{fig3} illustrates dependence $\eta_0(E;\nu)$. For $\eta_0\lesssim1$ it is well
described by Equation~(\ref{zeta_0}) with subtracted ``inner border'' value $\epsilon/\tilde v$, as determined by
condition~(\ref{zeta_min}). For large $\eta_0$, the exact dependence becomes unimportant for calculating $S(E)$, since the
exponential in Equation~(\ref{flux}) can be safely neglected.

Let us summarize the behavior of $S(E)$. At sufficiently high energies, the CR flux is not affected by turbulence and equal
to the free-streaming value,
\begin{equation}\label{flux0}
E>E_{\rm ex}:\qquad S_{\rm free}(E)=4\pi\langle\mu\rangle j_{\rm IS}(E).
\end{equation}
A continuous transition to the modulated flux occurs at $E=E_{\rm ex}(\nu)$, determined by Equation~(\ref{threshold}). For
smaller $E$, from Equations~(\ref{flux}) and (\ref{zeta_0}) we obtain the following general formula:
\begin{equation}\label{flux_final}
E<E_{\rm ex}:\qquad \frac{S(E)}{4\pi j_*}=\frac{\tilde E+1}{\sqrt{\tilde E(\tilde E+2)}}
\left(\frac{\epsilon \tilde j_{\rm IS}(E)}{1-e^{-\eta_0(E)}}\right),
\end{equation}
with diffusion depth
\begin{equation}\label{zeta_00}
\eta_0(E)=\sqrt{\tilde E(\tilde E+2)}\:\frac{\tilde j_{\rm IS}(E)}{2\nu}\,.
\end{equation}

\begin{figure*}\centering
\includegraphics[width=0.95\textwidth,clip=]{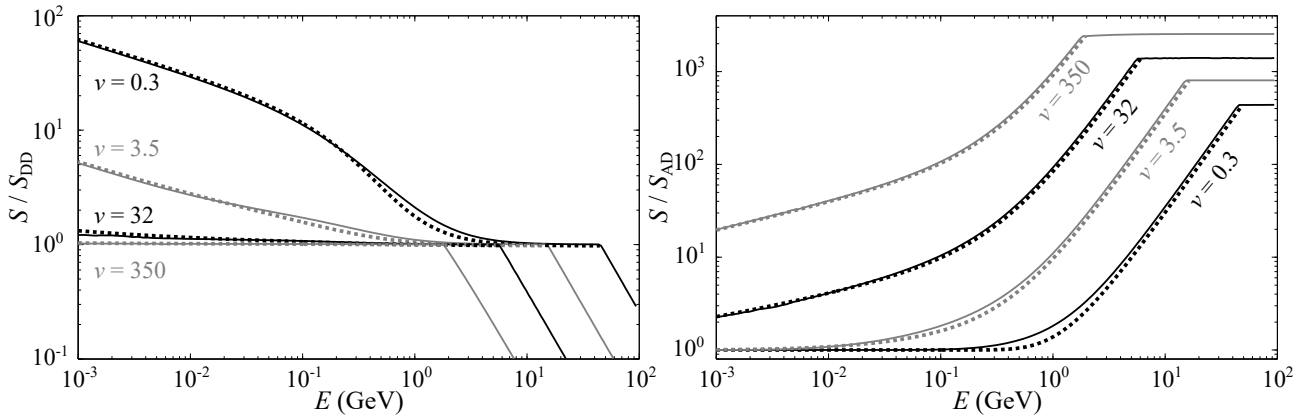}
\caption{Self-modulation of CRs. The solid lines show the numerically calculated energy dependence of CR flux, $S(E)$,
modulated by the self-generated turbulence, the dotted lines are analytical results obtained with
Equation~(\ref{flux_final}); the curves correspond to the values of $\nu$ in Figure~\ref{fig2}. To demonstrate the
asymptotic behavior at higher and lower energies, the left panel presents $S(E)$ divided by $S_{\rm DD}(E)$,
Equation~(\ref{flux1}), while in the right panel $S(E)$ is normalized by $S_{\rm AD}(E)$, Equation~(\ref{flux2}). }
\label{fig4}
\end{figure*}

For $\eta_0\lesssim1$, where the exponential in the denominator of Equation~(\ref{flux_final}) can be expanded, the
resulting leading term does not depend on $j_{\rm IS}(E)$. In this case we obtain ``diffusion-dominated'' flux,
\begin{equation}\label{flux1}
S_{\rm DD}(E)=\frac{\tilde E+1}{\tilde E+2}\left(\frac{8\pi\epsilon\nu j_*}{\tilde E}\right),
\end{equation}
where advection is unimportant and therefore its magnitude is governed by a balance of the excitation and damping in wave
equation~(\ref{W_SC}). This is the reason why it obeys a {\it universal} energy dependence, scaling as $\propto E^{-1}$ both
in the non-relativistic and ultra-relativistic limits (or, equivalently, as $\propto(pv)^{-1}$). Furthermore, from
Equations~(\ref{scale_nu}) and (\ref{scale_epsilon}) it follows that
\begin{equation}\label{flux_par}
S_{\rm DD}\propto m_{\rm g}n_{\rm g}\sqrt{\frac{n_{\rm i}}{m_{\rm i}}}\,,
\end{equation}
i.e., the flux does not depend on $j_*$ and thus is {\it solely} determined by the physical parameters of the envelope. We
want to emphasize that this expression can be deduced from a theoretical analysis by \citet[][]{skill76}, by substituting
their Equation~(6) into the second term of their Equation~(8).

At even lower energies, $\eta_0$ exceeds unity for smaller $\nu$, as evident from Figure~\ref{fig3}. Then advection
dominates and the flux tends to $v_{\rm A}N_{\rm IS}(E)$, which is
\begin{equation}\label{flux2}
S_{\rm AD}(E)= \frac{\tilde E+1}{\sqrt{\tilde E(\tilde E+2)}}\:4\pi\epsilon j_{\rm IS}(E).
\end{equation}
The analysis performed by \citet[][]{gabici15} corresponds to our case $\eta_0\sim1$, and therefore their conclusion that
the velocity of the CR flux penetrating into a cloud is of the order of $v_{\rm A}$ represents a crossover to the
advection-dominated flux.

Figure~\ref{fig4} shows the modulated CR flux obtained analytically, from Equation~(\ref{flux_final}) for IS
spectrum~(\ref{IS_spectrum}), and compared with the numerically calculated flux. One can see that the analytical results
provide a fairly accurate description of $S(E)$ in the whole energy range; only for very small $\nu$, a slight deviation
(about 50\%) is observed at intermediate energies, where $\eta_0(E)\sim1$ (as one can see from Figure~\ref{fig3}).

Both panels of the figure clearly demonstrate a transition from free streaming to the diffusive regime, occurring at
$E=E_{\rm ex}(\nu)$ and manifested by a kink at each curve.
In the left panel the curves are normalized by $S_{\rm DD}(E)$ and, hence, at $E<E_{\rm ex}$ they collapse into the
horizontal line at the unity level as long as $\eta_0(E)\lesssim1$ (for $E>E_{\rm ex}$ they approximately scale as $\propto
E\tilde j_{\rm IS}(E)/\nu^{3/4}$). In the right panel $S(E)$ is normalized by $S_{\rm AD}(E)$, and thus a crossover to the
advection-dominated flux occurs if the curves approach the unity level (for $E>E_{\rm ex}$ the curves tend to
$\epsilon^{-1}\propto\nu^{1/4}$). The crossover takes place only for small $\nu$, otherwise the flux remains
diffusion-dominated at all energies shown.

We point out that Equation~(\ref{flux_final}) is insensitive to the particular model of nonlinear wave cascade. As shown in
Section~\ref{approximate}, the cascade term in Equation~(\ref{W_SC}) is negligible for small $\eta_0$ (where $S\simeq S_{\rm
DD}$), whereas for large $\eta_0$ the CR flux tends to the advection asymptote $v_{\rm A}N_{\rm IS}$, i.e., the cascade term
may affect the flux only near the crossover point $\eta_0(E)\sim1$. This has been verified with numerical calculations
performed for the Kolmogorov cascade \citep[with $T_{\rm NL}$ taken from][]{ptus06}, indeed showing minor deviations from
the presented results in the crossover energy range.

\section{Effect of energy losses}
\label{realistic}

In the previous Section we derived intrinsic properties of the turbulent diffusion zone generated under idealized
conditions, where CRs propagate toward an absorbing wall, and the energy losses due to interaction with gas are unimportant.
This approach presumes the intrinsic spatial scale of the diffusion zone, $z_0(E)$, to be much smaller than the CR loss
length at a given energy. For realistic parameters of diffuse envelopes, the latter assumption is not always justified,
especially in the non-relativistic case.

For this reason, let us now move away from the initial assumption that CRs propagate freely through the envelope until they
reach the turbulent zone near the absorbing wall, to see what impact the gas losses may have on the diffusion and, most
importantly, how the flux self-modulation is affected by the losses.

\begin{figure}\centering
\includegraphics[width=0.9\columnwidth,clip=]{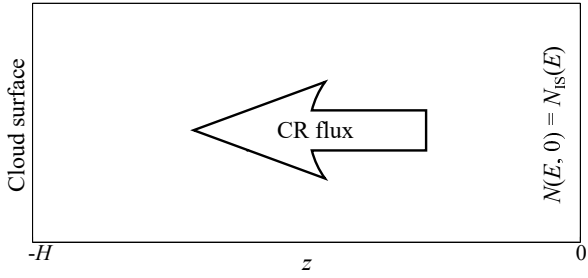}
\caption{Propagation of CRs in a low-density envelope with energy losses taken into account. The outer boundary of the
envelope (of size $H$) is now at $z=0$, with the same boundary condition as in Figure~\ref{fig1}.}
\label{fig5}
\end{figure}

The principal difference introduced to the problem by the gas losses is that the CR flux is no longer conserved, as follows
from Equation~(\ref{Diff_norm}). Therefore, the losses naturally generate a CR density gradient and, hence, stimulate wave
excitation across the whole envelope, starting from its outer boundary (whereas before the gradient was only present near
the absorbing wall). For this reason it is more convenient to analyze results in the frame of reference where $z=0$ is
located at the outer boundary, as shown in Figure~\ref{fig5}. Thus, now $|z_0|$ is referred to as the inner (``downstream'')
border of the diffusion zone and $|z_{\rm min}|~(<|z_0|)$ is the outer (``upstream'') border.

\subsection{Solution for the excitation-damping balance}
\label{inhomogeneous}

The general excitation criterion~(\ref{threshold}) does not depend on a particular problem setup and, hence, can also be
used when the losses are present. Turbulence sets in (and, as pointed out in Section~\ref{model}, the diffusive
approximation is thereby justified) when the excitation term on the rhs of Equation~(\ref{Wave_norm}) becomes equal to
damping. Furthermore, the role of the cascade term on the lhs remains largely negligible at $k\gtrsim k_{\rm ex}$: as we
demonstrate below in this Section, the condition of applicability of the excitation-damping balance is relaxed compared to
the loss-free case (where the cascade term can be neglected for $\eta_0\lesssim1$). Therefore, from
Equation~(\ref{Wave_norm}) we obtain
\begin{equation}\label{Wave_norm2}
\frac{\tilde D}{2\tilde k}\frac{\partial \tilde j}{\partial \tilde z}\simeq\nu.
\end{equation}
We see that $\tilde D\partial \tilde j/\partial \tilde z$, the diffusion part of flux~(\ref{smin_norm}), does not depend on
coordinates (for given $\nu$) and therefore does not contribute to transport equation~(\ref{Diff_norm}). The latter is then
reduced to
\begin{equation}\label{Diff_norm2}
\frac{\partial \tilde j}{\partial\tilde z}=-\frac{\partial}{\partial \tilde p}\left(\tilde L_{\rm g}\tilde j\right),
\end{equation}
giving the local CR spectrum, i.e., the advection part of flux~(\ref{smin_norm}).

Equation~(\ref{Diff_norm2}) has a general solution in $(p,z)$ space,
\begin{equation}\label{solution_gen}
\tilde L_{\rm g}(p)\tilde j(p,z)=\Phi\left(\tilde z-\int \frac{d\tilde p}{\tilde L_{\rm g}(\tilde p)}\right),
\end{equation}
where function $\Phi(x)$ is determined by the boundary condition $\tilde j(p,0)=\tilde j_{\rm IS}(p)$. To illustrate the
overall behavior and obtain useful closed-form expressions, let us again consider a power-law IS energy spectrum, $\tilde
j_{\rm IS}(E)=\tilde E^{-\alpha}$, and treat separately the non-relativistic and ultra-relativistic cases.

For $\tilde E\lesssim1$ the gas losses are dominated by ionization \citep[][]{hayakawa}. The loss function can be
approximated by $\tilde L_{\rm g}(E)\simeq A_{\rm ion}\tilde E^{-b}$, with the exponent in the range of $0\lesssim
b\lesssim1$. The solution resulting from Equation~(\ref{solution_gen}) is
\begin{equation}\label{solution1}
\tilde j(E,z)=\tilde j_{\rm IS}(E)\left(1+(2b+1)\frac{\tilde L_{\rm g}(E)}{\sqrt{2\tilde E}}|\tilde z|\right)
^{-\frac{2(b+\alpha)}{2b+1}}.
\end{equation}
The standard expression for non-relativistic ionization losses with $b=1$ is determined by \citep[][]{ginz79}
\begin{equation*}
A_{\rm ion}=\frac38\frac{m_{\rm e}}{m_{\rm p}}\frac{n_{\rm g}z_*\sigma_{\rm T}\ln\Lambda}{\epsilon}\,,
\end{equation*}
where $\Lambda$ is the argument of the Coulomb logarithm for the ionization losses (for hydrogen, $\Lambda\simeq20$),
$\sigma_{\rm T}=6.6\times10^{-25}$~cm$^{-2}$ is the Thomson cross section of electron, and $m_{\rm e}/m_{\rm p}=1/1836$ is
the electron-to-proton mass ratio.

In the relativistic case, the pion production occurring in proton-proton collisions above the threshold energy of
$\simeq280$~MeV is the main mechanism for the energy losses  \citep[][]{hayakawa}. The loss function can be approximated by
$\tilde L_{\rm g}(E)\simeq A_{\pi}\tilde E$, where \citep[][]{mann94}
\begin{equation*}
A_{\pi}=0.65\frac{n_{\rm g}z_*\sigma_{\pi}}{\epsilon}
\end{equation*}
is proportional to the effective cross section $\sigma_{\pi}\simeq3\times10^{-26}$~cm$^{-2}$ (neglecting a weak logarithmic
energy dependence). Then Equation~(\ref{solution_gen}) yields
\begin{equation}\label{solution2}
\tilde j(E,z)=\tilde j_{\rm IS}(E)e^{-(\alpha-1)A_{\pi}|\tilde z|}\,.
\end{equation}

The derived results also allow us to verify the (initially assumed) excitation-damping balance, Equation~(\ref{Wave_norm2}),
i.e., to identify conditions when the cascade term in Equation~(\ref{Wave_norm}) is negligible: Since the relative
contribution of the cascade term increases with $k$ (i.e., with decreasing $E$), it is sufficient to consider the
non-relativistic case. Substituting Equations~(\ref{solution1}) in Equation~(\ref{Wave_norm2}) and taking into account
Equation~(\ref{D_norm}) gives an estimate for $W(k)$, to be inserted in the lhs of Equation~(\ref{Wave_norm}). We obtain
that the latter is small compared to $\nu$ when $(A_{\rm ion}/\nu^2)\tilde E^{-(\alpha+1/2)}\lesssim1$, which can be
equivalently rewritten as $\eta_0\lesssim\tilde E\nu/A_{\rm ion}$ with $\eta_0$ from Equation~(\ref{zeta_0}). Comparing this
with condition $\eta_0\lesssim1$ for the loss-free case, we conclude that for $\tilde E\gtrsim A_{\rm ion}/\nu$
($\simeq10^{-4}\nu^{1/4}$ for the presented results, i.e., for all energies shown) the excitation-damping balance is indeed
more easily satisfied in the presence of losses.

\subsection{Onset of diffusion zone}
\label{zone_onset}

A condition of applicability of the diffusive regime is that the CR mean free path, $\sim D/v$, is smaller than the
characteristic spatial scale. In dimensionless form, the mean free path $\sim\epsilon\tilde D/ \tilde v$ should be smaller
than the relevant scale of the present problem, $\sim|\tilde z|$. By employing Equation~(\ref{Wave_norm2}), the condition is
reduced to
\begin{equation}\label{condition}
\frac{\tilde E+2}{\tilde E+1}\tilde E|\tilde z|\frac{\partial \tilde j}{\partial \tilde z}\gtrsim2\epsilon\nu,
\end{equation}
where $\tilde j(E,z)$ is a solution of Equation~(\ref{Diff_norm2}).

Equation~(\ref{condition}) is the {\it necessary} condition of applicability of the diffusive regime in the presence of
losses. For given $E$, its lhs is a function of $z$, whose maximum is of the order of $\sim\tilde E\tilde j_{\rm IS}(\tilde
E)$. Hence, for $\tilde j_{\rm IS}(E)=\tilde E^{-\alpha}$ condition~(\ref{condition}) essentially coincides with
criterion~(\ref{zone_eq}) of the diffusive regime, derived for the absorbing wall problem.

The sufficient applicability condition requires that the diffusion zone is formed within the envelope, i.e., that the outer
border $|z_{\rm min}(E)|$ at which inequality~(\ref{condition}) is first fulfilled is smaller than the envelope size $H$.
For the loss mechanisms discussed in Section~\ref{inhomogeneous}, we have
\begin{align}
\tilde E\lesssim1:&\qquad |\tilde z_{\rm min}|\sim\frac{\epsilon\nu}{(\alpha+1)A_{\rm ion}}\tilde E^{\alpha+1/2}\label{z_min1}\\
\intertext{and}
\tilde E\gtrsim1:&\qquad |\tilde z_{\rm min}|\sim\frac{\epsilon\nu}{(\alpha-1)A_{\pi}}\tilde E^{\alpha-1}.\label{z_min2}
\end{align}
Since $A_{\rm ion}/A_{\pi}\simeq7\times10^{-3}\ln\Lambda$ is practically a constant $\sim0.1$, a smooth crossover between
the two cases occurs at energy about a few tenths of GeV. With Equation~(\ref{z*}) we notice that in absolute units,
\begin{equation}\label{scaling_z2}
|z_{\rm min}|\propto \frac{B}{j_*m_{\rm i}}\,,
\end{equation}
the coordinate of the diffusion onset is proportional to $B$ and does not depend on $n_{\rm g}$ or $n_{\rm i}$. As regards
the dependence on $E$, it is determined by a particular IS energy spectrum. In Figure~\ref{fig6} (discussed in the next
Section), $|z_{\rm min}(E)|$ is the left border of the plotted diffusion zone, calculated for IS
spectrum~(\ref{IS_spectrum}); it scales approximately as $\propto E^{1.3}$
in the non-relativistic case.

Once requirement $|z_{\rm min}|\lesssim H$ is fulfilled and the diffusive regime operates, the dimensionless CR flux is
given by the corresponding expression in Equation~(\ref{smin_norm}), with $\tilde D\partial \tilde j/\partial \tilde
z=2\tilde k\nu$ and $\tilde j(E,z)$ from Equation~(\ref{Diff_norm2}). We see that the diffusion part of the modulated flux
dominates over the advection part when $2\tilde k\nu \gtrsim \tilde j$. This remarkably coincides with condition
$\eta_0\lesssim1$ of the diffusion-dominated flux for the loss-free case -- with the only difference that now $\eta_0$
should be evaluated not for $\tilde j_{\rm IS}(E)$ but for derived $\tilde j(E,z)$. Then the modulated flux (in absolute
units) is still given by Equation~(\ref{flux1}) obtained for the loss-free case; moreover, in the presence of losses,
$S_{\rm DD}(E)$ dominates over a broader range of parameters, since $\eta_0$ should be additionally multiplied by a factor
of $j/j_{\rm IS}\leq1$.

If advection dominates over diffusion, transport equation~(\ref{Diff_norm2}) still describes the advection part of
flux~(\ref{smin_norm}). In this case, the modulated flux is given by Equation~(\ref{flux2}) with $\tilde j_{\rm IS}(E)$
replaced by $\tilde j(E,z)$.

\subsection{CR flux}
\label{zone_onset2}

Summing up the above results, we conclude that the modulated CR flux in the presence of losses can be written as a simple
superposition of the diffusion and advection asymptotes: The diffusion flux is given by Equations~(\ref{flux1}), and the
advection flux is described by a modified Equation~(\ref{flux2}), with $\tilde j_{\rm IS}(E)$ replaced by solution $\tilde
j(E,z)$ of Equation~(\ref{Diff_norm2}). This yields
\begin{equation}\label{flux_final2}
S(E,z)\simeq S_{\rm DD}(E)\left(1+\sqrt{\tilde E(\tilde E+2)}\:\frac{\tilde j(E,z)}{2\nu}\right),
\end{equation}
where the relative magnitude of the advection flux is equal to the modified diffusion depth~(\ref{zeta_00}).

It is noteworthy that the sum of $S_{\rm DD}$ and $S_{\rm AD}$ not only provides the correct asymptotic behavior -- as
demonstrated below, Equation~(\ref{flux_final2}) also allows us to accurately describe a crossover between them. This can be
understood by bearing in mind a remark we made in the end of Section~\ref{inhomogeneous}: At higher energies, the losses
tend to extend the range of applicability of the excitation-damping balance, Equation~(\ref{Wave_norm2}), which directly
determines $S_{\rm DD}(E)$. Therefore, Equations~(\ref{flux1}) remains accurate where the crossover to advection
occurs.\footnote{We remind that in the loss-free case, the excitation-damping balance always breaks down at the crossover
point $\eta_0\sim 1$, see Sections~\ref{approximate} and \ref{CR_flux}.} Moreover, the losses generally reduce the relative
magnitude of the advection flux, so that the crossover may not take place at all.

\begin{figure}\centering
\includegraphics[width=\columnwidth,clip=]{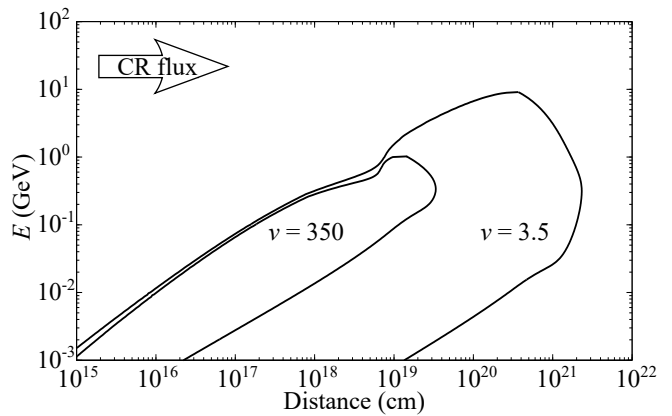}
\caption{Diffusion zone in the presence of losses (CR propagation is diffusive within the zone), plotted in $(E,|z|)$ plane
for two values of $\nu$. The outer (left) and inner (right) borders are $|z_{\rm min}(E)|$ and $|z_0(E)|$, respectively,
measured from the outer envelope boundary (see Figure~\ref{fig5}). Onset of the diffusive regime at a given energy requires
$|z_{\rm min}(E)|$ to be smaller than the envelope size $H$. Note that $|z_{\rm min}|$ does not depend on the gas or ion
densities (and hence on $\nu$), while $|z_0|$ rapidly decreases with $\nu$ (see Sections~\ref{zone_onset} and
\ref{zone_onset2}).} \label{fig6}
\end{figure}

>From Equations~(\ref{smin}) and (\ref{sfree}) it follows that the diffusive regime operates as long as the modulated flux,
approximately equal to $S_{\rm DD}(E)$, is smaller than the {\it local} free-streaming flux, which is proportional to
$\tilde j(E,z)$. Equation~(\ref{Diff_norm2}) suggests that this condition is violated at sufficiently large $|z|$, where
$\tilde j(E,z)$ becomes too small due to the losses. The corresponding inner border of the diffusion zone, $|z_0(E)|$, can
be directly obtained from excitation criterion~(\ref{threshold}) (written for given $E$) where, again, $\tilde j_{\rm
IS}(E)$ is replaced by $\tilde j(E,z)$,
\begin{equation}\label{threshold2}
\frac{\tilde E+2}{\tilde E+1}\tilde E\tilde j(E,z_0)=\frac{2\epsilon \nu}{\langle\mu\rangle}\,.
\end{equation}
Here, $\langle\mu\rangle$ is the average pitch angle of CRs for $|z|>|z_0(E)|$, which corresponds to a ``downstream''
free-streaming zone (see Appendix~\ref{<mu>}). Since the exact value of $\langle\mu\rangle\sim1$ is unimportant for the
presented analysis, for simplicity we keep the same notation as for the CR flux in the free-streaming zone~I.

The diffusion zone in the presence of losses is shown in Figure~\ref{fig6}, where the left border $|z_{\rm min}(E)|$ is
determined from condition~(\ref{condition}) and the right border $|z_0(E)|$ is derived from Equation~(\ref{threshold2}). The
overall shape of the zone and its qualitative change with $\nu$ are quite similar to what we see in Figure~\ref{fig2} for
the absorbing-wall case (we remind that distance $|z|$ in Figure~\ref{fig6} is measured in the negative direction). However,
$|z_{\rm min}|$ and $|z_0|$ are much larger than the respective spatial scales ($z_{\rm min}$ and $z_0$) in
Figure~\ref{fig2}. Also, Equation~(\ref{scaling_z2}) shows that $|z_{\rm min}|$ does not depend on $\nu$, i.e., the
diffusion zone shrinks due to a rapid decrease of $|z_0|$ with $\nu$,\footnote{In the presence of losses, the dependence of
$|z_0|$ on the physical parameters is different in the non-relativistic and relativistic cases, as one can see by
substituting Equations~(\ref{solution1}) and (\ref{solution2}) in Equation~(\ref{threshold2}).} while for the absorbing-wall
case both borders move toward each other as $\nu$ increases (see Equation~(\ref{scaling_z1})).

\begin{figure}\centering
\includegraphics[width=\columnwidth,clip=]{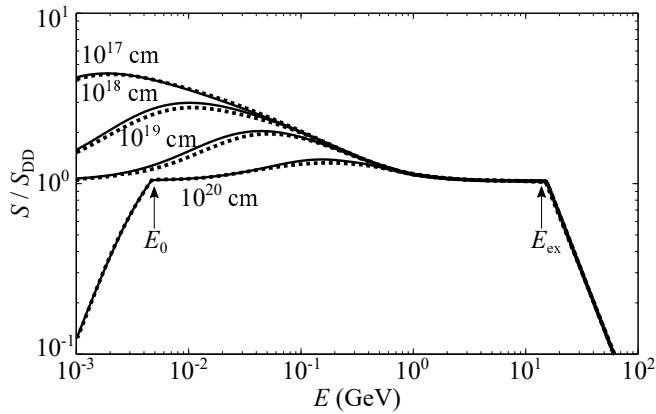}
\caption{Self-modulation of CRs in the presence of losses. Different curves depict the modulated flux $S(E,z)$ for
different distances $|z|$, as indicated; $S(E,z)$ is normalized by $S_{\rm DD}(E)$, as in the left panel of
Figure~\ref{fig4}. The solid lines are numerical calculations and the dotted lines are analytical results, both
corresponding to $\nu=3.5$: The diffusive regime at $E_0(\nu,z)<E<E_{\rm ex}(\nu)$ is described by
Equation~(\ref{flux_final2}), and the free-streaming regime induced by the losses at $E<E_0$ is represented by
Equation~(\ref{solution_gen2}). The matching energy $E_0$ for given $|z|$ (seen here only for $|z|=10^{20}$~cm) is obtained
by inverting $z_0(E)$.} \label{fig7}
\end{figure}

The free-streaming flux $S_{\rm free}(E,z)= 4\pi \langle\mu\rangle j(E,z)$ at $|z|>|z_0(E)|$ (as well as for $E>E_{\rm ex}$)
is determined by $j(E,z)$ which is a solution of transport equation~(\ref{Diff_norm}). A general form of the solution in
$(p,z)$ space is
\begin{equation}\label{solution_gen2}
\tilde L_{\rm g}(p)\tilde j(p,z)=\Phi\left(\tilde z-\langle\mu\rangle\int \frac{d\tilde p\:\tilde v}{\epsilon
\tilde L_{\rm g}(\tilde p)}\right),
\end{equation}
and the resulting $S_{\rm free}(E,z)$ has to be matched at $z=z_0(E)$ with Equation~(\ref{flux_final2}). Of course, the free
streaming regime is only realized when $|z_0(E)|<H$, otherwise the CR flux penetrating the cloud is directly given by
Equation~(\ref{flux_final2}).

The characteristic behavior of the modulated CR flux in the presence of losses is illustrated in Figure~\ref{fig7} for
$\nu=3.5$, again calculated for IS spectrum~(\ref{IS_spectrum}). One can see that the analytical curves obtained from
Equation~(\ref{flux_final2}) are in excellent agreement with the numerical results. The way how the losses modify the flux
is evident by comparing these curves with the corresponding loss-free curve plotted in the left panel of Figure~\ref{fig4}:
The flux is attenuated with the distance at lower energies, thus suppressing a crossover to the advection-dominated flux,
clearly seen in Figure~\ref{fig4} for $\nu=3.5$ (where the curve in the left panel steadily increases toward smaller $E$).
Furthermore, at $|z|>|z_0(E)|$ the losses induce a ``backward'' transition to the free-streaming regime, seen as the kink
for $|z|=10^{20}$~cm. For larger $\nu$ (not shown here), where the advection contribution is practically negligible, the
curves become almost horizontal in the diffusive regime and, hence, undistinguishable from those in Figure~\ref{fig4}. This
striking similarity is a manifestation of the universal behavior characterizing the diffusion-dominated flux $S_{\rm
DD}(E)$.

\section{Discussion and conclusions}
\label{discussion}

A comparison of results obtained in Sections~\ref{model} and \ref{realistic} demonstrates that, when calculating the
magnitude of the modulated CR flux, it is largely unimportant what leading mechanism -- absorbing wall or gas losses --
causes the self-modulation: Figure~\ref{fig6} suggests that in the presence of losses the condition of diffusion onset,
$|z_{\rm min}(E)|\lesssim H$, is usually fulfilled for non-relativistic CRs (assuming typical envelope size of 3--10~pc),
and hence they are modulated due to turbulence induced near the outer envelope boundary. For relativistic CRs losses are
typically unimportant at a scale of the envelope, and their self-modulation occurs near the absorbing cloud wall; according
to Figure~\ref{fig2}, the respective condition $z_{\rm min}(E)<H$ is well satisfied. Nevertheless,
the resulting CR flux remains {\it universal at all energies} below $E_{\rm ex}$ -- it is described by the
diffusion-dominated asymptote $S_{\rm DD}(E)$, Equation~(\ref{flux1}). Figures~\ref{fig4} and \ref{fig7} indicate that the
effect of advection, causing a deviation from this dependence, only becomes significant if $\nu\lesssim10$ (according to
Equation~(\ref{scale_nu}), the corresponding gas density in the envelope typically must be well below $\sim100$~cm$^{-3}$).

Of course, the gas losses can destroy universality of the energy spectrum for low-energy CRs penetrating into the cloud:
Figure~\ref{fig6} shows that, at lower energies and for sufficiently large $\nu$~($\gtrsim100$), the right border of the
diffusion zone $|z_0(E)|$ becomes smaller than typical $H$. As discussed in Section~\ref{realistic}, the further
free-streaming propagation of such CRs in the envelope is described by Equation~(\ref{solution_gen2}), and their flux is
proportional to the local spectrum $j(E,z)$. If the remaining distance $H-|z_0(E)|$ exceeds the integral term in the
parentheses (multiplied by $z_*$), the attenuation modifies the universal spectrum of $S_{\rm DD}(E)$ before CRs reach the
cloud.

The presented results allow us to address several important questions regarding interaction of CRs with molecular clouds,
and draw the following major conclusions:
\begin{enumerate}
\item {\it Dimensionless numbers.} Generic features of CR propagation in low-density envelopes are completely determined
    by two dimensionless numbers: gas damping rate $\nu$, Equation~(\ref{nu}), which governs the diffusive transport
    regime (due to the self-generated MHD turbulence), and small parameter $\epsilon$, Equation~(\ref{epsilon}), which
    controls a transition between the diffusive regime and a free streaming of CRs (where the turbulence is
    unimportant).
\item {\it Diffusive propagation.} The turbulence generated by CRs in the envelope affects their transport at energies
    below the excitation threshold $E_{\rm ex}$, Equation~(\ref{threshold}), which is a function of the product
    $\epsilon\nu$. As a result, the CR flux becomes self-modulated before penetrating into the cloud -- it changes from
    a free-streaming flux, determined by given IS energy spectrum $j_{\rm IS}(E)$, to the universal diffusion-dominated
    flux $S_{\rm DD}(E)$, scaling as $\propto E^{-1}$ both in the non-relativistic and ultra-relativistic limits. The
    locations of the diffusion zones (regions of the diffusive propagation) in the envelope are determined by the
    leading mechanism of self-modulation for given $E<E_{\rm ex}$: The zone can either be formed near the inner boundary
    (for higher-energy CRs, whose propagation is unaffected by the gas losses) or near the outer boundary (for lower
    energies, where the losses are essential).
\item {\it Wave losses.} In Section~\ref{WL} we showed that taking into account the wave losses basically leads to a
    renormalization of the advection flux $S_{\rm AD}(E)$, Equation~(\ref{flux2}). Since a contribution of $S_{\rm AD}$
    to the modulated CR flux is significant only for relatively small $\nu$, the effect of wave losses can be
    practically always neglected.
\item {\it Important physical parameters.} The excitation threshold $E_{\rm ex}(\epsilon\nu)$ does not depend on the
    magnetic field $B$; it is a function of the physical parameters of the envelope as well as of the magnitude and the
    form of $j_{\rm IS}(E)$. One of our key findings is that the universal flux $S_{\rm DD}(E)$ is insensitive to the
    particular model of nonlinear wave cascade, depends neither on $B$ nor on $j_{\rm IS}(E)$, and thus is only
    determined by densities and masses of the neutral and ionized species in the envelope, Equation~(\ref{flux_par}).
\item {\it Magnitude of the self-modulation.} The CR modulation due to self-generated turbulence is conveniently
    characterized by the flux ratio
\begin{equation*}
\frac{S_{\rm DD}(E)}{S_{\rm free}(E)}\simeq\frac{\epsilon\nu}{\tilde E\tilde j_{\rm IS}(E)}\,,
\end{equation*}
    determined by Equations~(\ref{flux0}) and (\ref{flux1}). For IS spectra analogous to that of
    Equation~(\ref{IS_spectrum}), the product $\tilde E\tilde j_{\rm IS}(E)$ achieves a broad maximum ($\sim1$) at
    $E\sim100$~MeV. Therefore, the strongest modulation occurs at these energies, where the reduction is
    $\sim\epsilon\nu$; for typical envelopes, the flux can decrease by up to two orders of magnitude.
\end{enumerate}

The conclusion that the CR flux penetrating into denser cloud regions has a universal energy dependence at $E<E_{\rm ex}$,
solely determined by the physical parameters of the envelope, is of substantial general interest and importance. One of the
reasons is that gamma-ray emission, measured from molecular clouds at different distances from the Galactic Center
\citep[see, e.g.,][]{digel,yang1,tib15}, is considered to provide information about the global distribution of CRs in the
Galaxy \citep[see e.g.][]{ahar01,casa10}. The derived spatial distribution of Galactic CRs is then interpreted as a result
of global-scale CR propagation and used as an input for models of the CR origin \citep[see,
e.g.,][]{Bloe2,Breit,galprop,recc16a}. Thus, the fact that the modulated flux is independent of the spectrum of Galactic CRs
may have profound implications for such analysis.

Also, observations indicate that the central regions of the Galactic Disk are enhanced by molecular hydrogen in the form of
very dense molecular clouds and diffuse gas \citep{oka05}. The latter occupies about 30\% of the volume of the central
molecular zone, and therefore the overall effect of the local self-modulation, which we predict to occur in these diffuse
regions, can be significant. For example, the spectrum of CR protons deduced by \citet{acero16} and \citet{yang3} from the
Fermi data for the inner Galaxy is harder than that in the outer Galaxy, and one can speculate that this may be due to the
local self-modulation.

The self-modulation of a CR flux can be important for many other fundamental problems. In particular, this could cause the
substantial reduction of CR ionization rates observed within dense molecular clouds \citep[e.g.,][]{caselli98},
significantly lower than those measured toward diffuse clouds \citep[][]{indri12}. We note that drops in the amount of CR
flux, and the consequent drop in the CR ionization rate within (UV-)dark clouds, affect physical parameters crucial for the
dynamical evolution of dense clouds: the ionization fraction, which controls the coupling between gas and magnetic fields,
thus regulating star formation \citep[e.g.][]{mckee89}; the gas temperature, which determines the thermal pressure,
particularly important at the scales of dense cloud cores \citep[e.g.,][]{fuller92,keto08} where stars form; internal MHD
turbulence in molecular clouds, which could contribute to the observed magnetic and virial equilibrium and thus to the cloud
dynamics and evolution \citep[e.g.,][]{myers88,goodman98,caselli02}. Last but not least, changes in the CR flux can
significantly affect the chemistry, as gas-phase processes in dark clouds are dominated by ion-molecules reactions with
rates depending on the ionization fraction \citep[][]{herbst73}, while surface chemistry can be modified by CRs directly
(via the impulsive spot heating) or indirectly (via the UV-photons generated by the fluorescence of H$_2$ molecules).

Self-consistent numerical simulations of dynamically and chemically evolving magnetized interstellar clouds (with a proper
treatment of CR propagation inclusive of their self-modulation and MHD turbulence generation) are needed to quantify our
predictions for case-specific clouds within our Milky Way and external galaxies, as well as to test our theory against
observations.

\section*{Acknowledgements}

The authors are grateful to Andy Strong for reading the manuscript and giving useful comments. VAD and DOC are supported in
parts by the grant RFBR 18-02-00075. DOC is supported in parts by foundation for the advancement of theoretical physics
``BASIS''. PC acknowledges support from the European Research Council (ERC) Advanced Grant PALs 320620. CMK is supported in
part by the ROC Ministry of Science and Technology grants MOST 104-2923-M-008-001-MY3 and MOST 105-2112-M-008-011-MY3. KSC
is supported by the GRF Grant under HKU 17310916.

\appendix

\section{Appendix A\\ Average pitch angle in the free-streaming regime}\label{<mu>}

Different transport zones are sketched in Figure~\ref{figA1}. For certainty, the zones are illustrated for the
absorbing-wall setup (distance~$=z$, see Figure~\ref{fig1}); the results are then readily applied to the setup with losses
(distance~$=|z|$, see Figure~\ref{fig5}). One can identify three free-streaming zones:

\begin{figure*}[ht]\centering
\includegraphics[width=0.4\textwidth,clip=]{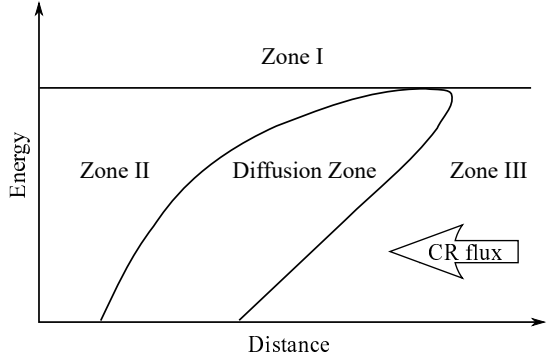}
\caption{Sketch of the transport zones in energy-distance plane, representing the absorbing-wall setup. For the setup with
losses (where the flux is directed to the right), the labels ``zone II'' and ``zone III'' should be swapped.} \label{figA1}
\end{figure*}

In zone I, corresponding to $E>E_{\rm ex}(\nu)$, CRs propagate across the envelope without experiencing scattering at any
distance. The value of $\langle\mu\rangle$ in this case depends on mechanisms governing modification of isotropic IS
spectrum $j_{\rm IS}(E)$ upon its entering into the envelope. (Since the strengths of the magnetic field inside and outside
the envelopes are about the same, it is reasonable to assume that the magnetic field lines enter into the envelope without
significant distortions.) Let us denote the spectrum formed upon entering as $j_{\rm IS}^*(E,\mu)$ with $\mu>0$. Then the
average pitch angle, which determines free-streaming flux $S_{\rm free}(E)$ in Equation~(\ref{sfree}), is readily obtained:
\begin{equation}\label{<mu>eq}
\langle\mu\rangle=\int_{0}^{1}d\mu\:\mu \frac{j_{\rm IS}^*(E,\mu)}{{j_{\rm IS}(E)}}\,.
\end{equation}
The exact form of $j_{\rm IS}^*(E,\mu)$ depends on unknown details of entering, but one can generally conclude that the
resulting value of $\langle\mu\rangle$ is of the order of a few tenths. For instance, if $j_{\rm IS}^*(E,\mu)$ is simply a
hemisphere $\mu>0$ of $j_{\rm IS}(E)$, then $\langle\mu\rangle=1/2$ (which corresponds to a well-known expression for a
free-streaming flux through a flat surface). Generally, $\langle\mu\rangle$ may be a function of $E$.

Zone II is located ``downstream'' from the diffusion zone. The value of $\langle\mu\rangle$ is determined by modification of
a local quasi-isotropic CR spectrum $j(E)$ leaving the diffusion zone. While details of this process may be different from
those controlling $\langle\mu\rangle$ in zone I, one can still employ Equation~(\ref{<mu>eq}) with $j_{\rm IS}(E)$ replaced
by $j(E)$. Using exactly the same line of arguments as before, we conclude that $\langle\mu\rangle$ in zone II should be
about that in zone I.

Zone III ``upstream'' from the diffusion zone is unimportant for our analysis. For $E\ll E_{\rm ex}$, the flux propagating
further toward the cloud the strongly modulated, i.e., the incident IS flux is almost entirely reflected back from the
diffusion zone. Therefore, the value of $\langle\mu\rangle$ in zone III is very small, tending to $\sim v_{\rm A}/v$ when
advection part of the (modulated) flux in Equation~(\ref{smin}) dominates over the diffusion part.

In the presence of losses, the ``upstream'' (``downstream'') zone corresponds to smaller (larger) distances (see
Figure~\ref{fig5}). Essentially, in this case we only need to swap zones II and III in the shown sketch.

\section{Appendix B\\ Numerical solution of the governing equations}\label{numerical}

Numerical results are deduced from the steady-state solution of time-dependent dimensionless Equations~(\ref{Diff_norm}) and
(\ref{Wave_norm}), obtained by adding terms $-\partial\tilde j/\partial\tilde t$ and $(2\tilde W)^{-1}\partial\tilde
W/\partial\tilde t$, respectively. Dimensionless time $\tilde t=t/t_*$ is determined by $t_*$ whose value is dictated by the
used normalization. We employ an explicit finite difference method, which has straightforward implementation and reasonable
convergence for our parameters.

To include the limitations on the CR flux velocity, we split this method into two steps: First, we evaluate the flux from
\begin{equation}\label{eq:SD_numerical}
\begin{array}{l}
{\displaystyle(\tilde{S}_{\rm diff})_{i,l}=\tilde{D}_{i,l}\frac{\tilde{j}_{i+1,l}
-\tilde{j}_{i,l}}{\tilde{z}_{i+1}-\tilde{z}_{i}}+\tilde{j}_{i+1,l}}\,,\vspace{.2cm}\\
\tilde{S}_{i,l} = \mbox{sign}\left\{(\tilde{S}_{\rm diff})_{i,l}\right\} \times \min \left\{|(\tilde{S}_{\rm diff})_{i,l}|,\:
(\tilde{S}_{\rm free})_{i+1,l}\right\},
\end{array}
\end{equation}
and then calculate the evolution of the CR energy spectrum $\tilde j_{i,l}$. Here indices $i$ and $l$ represent
discretization of the spatial coordinate and energy, respectively.

In fact, $(\tilde{S}_{\rm diff})_{i,l}$ in Equation~(\ref{eq:SD_numerical}) is evaluated at an intermediate grid point, for
which chose midpoint $\tilde z_{i+\frac12}=\frac12(\tilde z_i+\tilde z_{i+1})$. Therefore, also the diffusion coefficient
$\tilde{D}_{i,l}$ and the density of MHD waves $\tilde{W}_{i,l}$ are calculated at $z_{i+\frac12}$. However, for brevity we
omit $\frac12$ in the spatial index, keeping in mind that all these parameters actually correspond to the midpoint. Thus, a
discrete equation for the energy spectrum is written as
\begin{equation*}
\frac{\tilde{j}_{i,l}(t+\Delta t)-\tilde{j}_{i,l}(t)}{\Delta t}
=2\frac{\tilde{S}_{i,l}-\tilde{S}_{i-1,l}}{\tilde{z}_{i+1}-\tilde{z}_{i-1}}
+\frac{(\tilde{L}_g)_{i,l+1}\tilde{j}_{i,l+1}-(\tilde{L}_g)_{i,l}\tilde{j}_{i,l}}{\tilde{p}_{l+1}-\tilde{p}_{l}}\,,
\end{equation*}
where relation $\tilde{z}_{i+\frac12} - \tilde{z}_{i-\frac12} = \frac12(\tilde z_{i+1} - \tilde z_{i-1})$ is taken into
account. For small values of the diffusion coefficient, this becomes a standard explicit scheme for the heat transport
equation with central difference, otherwise it transforms into an upwind scheme.

The evolution of density of the MHD waves is performed in a similar way. We have verified that results do not practically
change when the advection wave transport, described by the first term on the lhs of Equation~(\ref{Wave_Eq}), is taken into
account. This allows us to omit this term and use the following upwind scheme:
\begin{equation*}
\frac{\tilde{W}_{i,l}(t+\Delta t)-\tilde{W}_{i,l}(t)}{\Delta t}
+\frac{\tilde{k}^{3}_{l+1}\tilde{W}^2_{i,l+1}-\tilde{k}^{3}_l \tilde{W}^2_{i,l}}{\tilde{k}_{l+1}-\tilde{k}_l}
=2(\tilde{\Gamma}_{i,l} - \tilde{\nu})\tilde{W}_{i,l}\,,
\end{equation*}
where $\tilde{\Gamma}_{i,l} = \tilde{S}_{i,l}/(2\tilde{k}_l)$; for $2(\tilde{\Gamma}_{i,l} - \tilde{\nu})\Delta t \ll 1$,
the last term is replaced by $2(\tilde{\Gamma}_{i,l} - \tilde{\nu})\tilde{W}_{i,l}(t+\Delta t)$. To simplify the problem, we
utilize the same grid for $\tilde j$ and $\tilde W$, and therefore $\tilde{p}_l$ and $\tilde{k}_l$ are related through the
resonance condition.

Boundary conditions for the above equations are:
\begin{eqnarray}
&&\tilde{j}_{1,l} = 0, \nonumber \\
&&\tilde{j}_{\mathcal I,l} = (\tilde{j}_{\rm IS})_l\,, \nonumber \\
&&\tilde{j}_{i,\mathcal L} = 0, \nonumber \\
&&\tilde{W}_{i,\mathcal L} = 0,\nonumber
\end{eqnarray}
where $\mathcal I$ and $\mathcal L$ denote the number of points on $z$ and $E$ (or $k$) axes, respectively.

In order to accelerate the relaxation process, we assume that CRs are uniformly distributed at the initial moment, i.e.,
$\tilde{j}(t=0) = \tilde{j}_{\rm IS}$. As for the waves, we introduce a certain ``zero-level'' turbulence at the initial
moment, and also ensure that $W$ never decreases below that level during its evolution. The choice of zero-level turbulence
is dictated by two conditions: first, this should not affect CR propagation; second, this should be large enough for a fast
convergence. The first condition is satisfied if the corresponding diffusion coefficient is $\sim\Theta vH$ with $\Theta \gg
1$, whereas the convergence time logarithmically depends on $\Theta$. Hence, a reasonably fast convergence can be archived
for a wide range of $\Theta$, for our calculations $\Theta = 10^{10}$ was chosen.

The energy loss function $L_{\rm g}(E)$ is calculated as a sum of the ionization and pion production terms. Ionization
losses, essential for non-relativistic protons, are taken from PSTAR NIST database \citep{star}, while for losses due to the
pion production we employ the expression proposed by~\citet{mann94}.

\section{Appendix C\\ Expansion of the wave spectrum in series of $z$}\label{linear_W}

\begin{figure*}[ht]\centering
\includegraphics[width=0.45\textwidth,clip=]{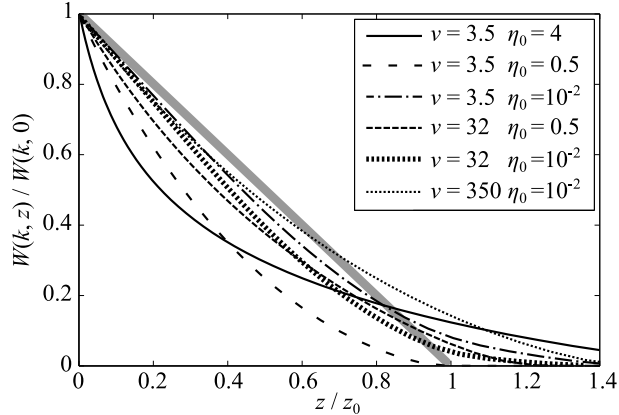}
\caption{Normalized wave spectrum $W(k,z)$, calculated numerically for different values of $\nu$ and $\eta_0$ (see the
legend in the inset). The analytical approximation~(\ref{field}) is the solid grey line. The coordinate is normalized by the
analytical $z_0(k)$, derived from solution~(\ref{w'}).} \label{figA2}
\end{figure*}

In Figure~\ref{figA2} we plot the wave spectrum $W(k,z)$ calculated numerically from Equation~(\ref{W_SC}).

One can see that, when $\nu\gg1$ and $\eta_0\lesssim1$, Equation~(\ref{field}) reasonably approximates the numerical results
except for a region near $z\simeq z_0$, where $W$ is relatively small. If needed, a quadratic term $\propto z^2$ can be
included in Equation~(\ref{field}) to further improve the agreement with the numerical results. For relatively small values
of $\nu$ and $\eta_0\gtrsim1$ the linear expansion fails to describe the results properly.

\end{document}